\documentclass{aa}  

\usepackage[T1]{fontenc}
\usepackage{ae,aecompl}

\usepackage{amssymb}	% Extra maths symbols
\usepackage{amsmath}
\usepackage{graphicx}
\usepackage[usenames,dvipsnames]{xcolor}
\usepackage{bm} 
\usepackage{url}

\usepackage[caption = false]{subfig}
\usepackage{float}   

\usepackage{xspace}
\usepackage{cancel}
\usepackage[normalem]{ulem}

\usepackage{hyperref}
 \hypersetup{
     colorlinks=true,
     linkcolor=blue,
     filecolor=blue,
     citecolor = teal,      
     urlcolor=olive,
 }

\newcommand{\change}[1]{{\textcolor{black}{#1}}}

\newcommand{\MpcOh}{ \,  \mathrm{Mpc}  \, h^{-1} }

\newcommand{\nn}{ \nonumber }

\usepackage{scalerel}

\newcommand{\beq}{\begin{equation}}
\newcommand{\eeq}{\end{equation}}
\newcommand{\beqa}{\begin{eqnarray}}
\newcommand{\eeqa}{\end{eqnarray}}

\begin{document}

%\title{ Combination of the self-calibration and clustering-$z$ information in the true-$z$ distribution inference }
%\title{ Joint inference of the true-$z$ distribution by   \\ the self-calibration and clustering-$z$ methods }
%\title{  Optimizing Redshift Distribution Inference through Joint Self-Calibration and Clustering-Redshift Strategies }
%\title{  Optimizing Redshift Distribution Inference through Joint Self-Calibration and Clustering-Redshift Synergy }
\title{  Optimizing redshift distribution inference through joint self-calibration and clustering-redshift synergy }
%Joint inference of the true redshift distribution by   \\ the self-calibration and clustering-redshift methods 
\titlerunning{ Joint inference by SC+CZ }

%% \subtitle{A Path to Observing the Turnaround Radius}

   \author{Weilun Zheng
         \inst{1}\fnmsep\inst{2} 
         \and
           Kwan Chuen Chan\thanks{ \email{chankc@mail.sysu.edu.cn (KCC)} }
         \inst{1}\fnmsep\inst{2}
         \and 
         Haojie Xu
         \inst{3}\fnmsep\inst{4}\fnmsep\inst{5}
         \and
         Le Zhang
         \inst{1}\fnmsep\inst{2}\fnmsep\inst{6}
         \and
         Ruiyu Song
         \inst{1}\fnmsep\inst{2}
          }
    \authorrunning{ Zheng  \&  Chan, et al}

   \institute{School of Physics and Astronomy, Sun Yat-Sen University, 2 Daxue Road, Tangjia, Zhuhai 519082, China 
        \and
        CSST Science Center for the Guangdong-Hongkong-Macau Greater Bay Area, SYSU, Zhuhai 519082, China
        \and
        Shanghai Astronomical Observatory, Chinese Academy of Sciences, Nandan Road 80, Shanghai 200240, China
        \and
        Department of Astronomy, Shanghai Jiao Tong University, Shanghai 200240, China
        \and 
        Key Laboratory for Particle Astrophysics and Cosmology (MOE)/Shanghai Key Laboratory for Particle Physics and Cosmology, China
        \and
        Peng Cheng Laboratory, No.2, Xingke 1st Street, Shenzhen 518000, China
   }

   \date{}

%\abstract{}{}{}{}{} 
% 5 {} token are mandatory
 
\abstract  % no empty line in between {}, except comments
 %% % context heading (optional)
{  Accurately characterizing the true redshift (true-$z$) distribution of a photometric redshift (photo-$z$) sample is critical for cosmological analyses in imaging surveys. Clustering-based techniques, which include clustering-redshift (CZ)  and self-calibration (SC) methods—depending on whether external spectroscopic data are used—offer powerful tools for this purpose.}
{In this study, we explore the joint inference of the true-$z$ distribution by combining SC and CZ (denoted as SC+CZ).}
{We derived simple multiplicative update rules to perform the joint inference. By incorporating appropriate error weighting and an additional weighting function, our method shows significant improvement over previous algorithms. We validated our approach using a DES Y3 mock catalog.}
{The true-$z$ distribution estimated through the combined SC+CZ method is generally more accurate than using SC or CZ alone.  To account for the different constraining powers of these methods, we assigned distinct weights to the SC and CZ contributions. The optimal weights, which minimize the distribution error, depend on the relative constraining strength of the SC and CZ data. Specifically, for a spectroscopic redshift sample that \change{amounts to} 1\% of the photo-$z$ sample, the optimal combination reduces the total error by 20\% (40\%) compared to using CZ (SC) alone, and it keeps the bias in mean redshift [$\Delta \bar{z} / (1 + z) $] at the level of 0.003.   Furthermore, when CZ data are only available in the low-$z$ range  and the high-$z$ range relies solely on SC data, SC+CZ enables consistent estimation of the true-$z$ distribution across the entire redshift range.}
{Our findings demonstrate that SC+CZ is an effective tool for constraining the true-$z$ distribution, paving the way for clustering-based methods to be applied at $z\gtrsim 1$.  }% , and holding promise for unbiased true-$z$ estimates within the range of .}

\keywords{cosmology: observations – gravitational lensing: weak – galaxies: photometry – surveys }

   \maketitle
%
%-------------------------------------------------------------------

\section{Introduction}
\label{introduction}

Wide-area imaging surveys provide powerful cosmological probes to constrain cosmology. Weak lensing is a prime example \citep{BartelmannSchneider2001,Heymans_etal2013,Hildebrandt_etal2017,Troxel_eltal2018,Asgari_etal2021,Hikage_etal2019,Amon_etal2022,Secco_etal2022,Li_etal2023,Dalal_etal2023}. In particular, cosmic shear is often bundled in 3$\times$2 point analysis, which includes cosmic shear, galaxy-galaxy lensing, and galaxy clustering \citep{DESY1_3x2pt, DESY3_3x2pt, Heymans_etal2021, Miyatake_etal2023,Sugiyama_etal2023}. These analyses offer a strong constraint on $S_8$, and the tightest uncertainty level ($\sim 2 \%$) is already comparable to that from the Planck \change{ Satellite's study of the cosmic microwave background } \citep{Planck2020}. Because  the lensing results are consistently lower than \change{ those from } the cosmic microwave background, there is heated discussion regarding the $S_8$ tension. Moreover, imaging surveys also enable measurements of the transverse baryon acoustic oscillations (BAO) \citep{Padmanabhan_etal2007,EstradaSefusattiFrieman2009, Hutsi2010, Seo_etal2012, Carnero_etal2012, deSimoni_etal2013, Abbott:2017wcz, DES:2021esc, Chan_xip2022, DESY6_BAO,Song_etal2024}.  The latest transverse BAO measurements, such as \cite{DESY6_BAO}, are yielding \change{constraints} competitive  with those from spectroscopic surveys \change{\citep{DESI_BAOmeasurement}}. Upcoming stage IV  surveys, such as from the  Rubin Observatory Legacy Survey of Space and Time (LSST) \citep{LSST_2019,LSST_DESC_2018}, Euclid \citep{Euclid_2011}, Chinese Space Station Telescope (CSST) \citep{Zhan_2011,Gong_etal2019}, and Roman Space Telescope (WFIRST) \citep{WFIRST_2015,WFIRST_2021}, are expected to deliver even more enormous photometric data and hence more exquisite results.

In imaging surveys, the redshifts (photo-$z$'s) are derived from the photometry information measured from a few broadband filters. The template fitting and training methods are commonly used to infer the photo-$z$'s of the galaxies (see \citet{Salvato_etal2019,NewmanGruen_2022} for a review). \change{The} template fitting method [e.g.~\citet{Arnouts_1999, Bolzonella_etal2000, Benitez_2000, Ilbert_etal2006}] fits a model derived from known spectral energy distribution SED) templates, which includes the photo-$z$ as a fitting parameter, to the color or magnitude data. The prior information can also be included in the fitting.  This method is limited by the accuracy and representativeness \change{of} the templates and  the accuracy of the prior information.  Training methods often make use of tools from machine learning \citep{CollisterLahav_2004, Sadeh:2015lsa, DeVicente:2015kyp, Zhou_etal2021, LiNapolitano_etal2022}.  A  spectroscopic redshift (spec-$z$) sample is required to train the machine learning algorithm, so its accuracy depends on the abundance and the completeness or representativity of the spec-$z$ sample.

\change{  In cosmological applications, accurate true redshift (true-$z$) distribution of the photo-$z$ sample must be obtained to avoid biasing the cosmological results.  Using a weighted spec-$z$ sample is a viable approach. }  The effectiveness of this method hinges on the  availability of the spec-$z$ sample or sometimes a high-quality photo-$z$ sample with numerous photo-$z$ bands.   The weighted spec-$z$ sample can be constructed using \change{a} $k$-nearest neighbor search \citep{Lima_etal2008, Cunha_etal2009, Bonnett_etal2016} or a self-organizing map \citep{KindBrunner_2014,Masters_etal2015,Buchs_etal2019,Wright_etal2020,Campos_etal2024}.

The clustering-based methods provide an independent way to calibrate the true-$z$ distribution of the photo-$z$ sample. Unlike photometry-based methods, this approach uses the clustering information, which can be traced back to gravity. Depending on the utilization of the external spec-$z$ sample, it can be further categorized into the clustering-$z$ (hereafter CZ)  and self-calibration (SC) methods.  In CZ, the true-$z$ distribution is determined by cross correlating the photo-$z$ sample with an external spec-$z$ sample \citep{Newman_2008, MatthewsNewman_2010, McQuinnWhite_2013, Menard_etal2013, Schmidt_etal2013, Morrison_etal2017, vandenBusch_etal2020}. This requires the spec-$z$ sample to overlap with the photo-$z$ one spatially, but the spec-$z$ sample does not need to be representative.   CZ has been routinely used to calibrate the true-$z$ distribution in real data [e.g.,~\cite{Gatti_etal2018, Gatti_etal2022, Cawthon_etal2022, Hildebrandt_etal2021, Rau_etal2023}].  On the other hand, the SC method relies solely on the clustering information, \change{both the auto and cross bin correlation function}, of the photometric sample itself. \citep{Schneider_etal2006, Zhang_etal2010, Benjamine_etal2010, Zhang_etal2017, Peng_etal2022, XU2023}.  Although SC has also been used for weak lensing \citep{Benjamine_etal2013} and BAO \citep{Song_etal2024} measurements, it is less frequently used than CZ. This may be because the redshift range explored in current surveys are still relatively low ($z\lesssim 1$), and hence the spec-$z$ sample is still sufficient for the calibration purpose.  However, even in present surveys, the high redshift portion of the sample cannot be calibrated using CZ due to the absence of spec-$z$ galaxies at high redshifts [e.g., \citet{Rau_etal2023, DESY6_BAO}]. Consequently, we expect the SC method to play a more prominent role in upcoming surveys.

In this paper we explore combining the information of CZ and SC to simultaneously constrain the true-$z$ distribution of a photo-$z$ sample.  As far as we know, this is the first time that these two methods have been jointly applied to constrain the true-$z$ distribution in realistic mock data [see the Fisher forecast in \cite{McQuinnWhite_2013}].  We anticipate that this synergy in redshift calibration is particularly fruitful in the high redshift regime, where the spec-$z$ sample is scarce.   The rest of the paper is organized as follows. We first review the SC and CZ formalism in Sec.~\ref{sec:formalism}  and then present the algorithm to jointly solve the SC and CZ equations in Sec.~\ref{sec:numerical_sol}. We test our method using a DES Y3 mock catalog in Sec.~\ref{sec:mock_test}.  In particular, we contrast the true-$z$ inference results from SC, CZ, and SC +CZ, and we demonstrate that  our improved algorithm is superior to the old algorithm. Moreover, we study the scenarios when the number of spec-$z$ bins is equal to and less than the number of photo-$z$ bins respectively, and show that SC+CZ can effectively extend the clustering-based method to a higher redshift. We conclude in Sec.~\ref{sec:conclusions}. In Appendix    
\ref{Appendix:Derivation_update_rule}, we present the derivation of the update rules. We investigate the impact of outlying spec-$z$ galaxies in Appendix \ref{sec:Impact_OutlyingGal} and the impact of setting the negative measurements to a tiny positive value in Appendix \ref{sec:Impact_NoNegative}.

\section{ The joint calibration method  }

\label{sec:formalism_methodology}  

\subsection{Derivation of the SC and CZ equations }
\label{sec:formalism}

In this subsection, we show how the correlation function of the photo-$z$ sample and the cross correlation function between the  photo-$z$  and spec-$z$ sample are related to the true-$z$ distribution.  Our convention follows that of \cite{Song_etal2024}, \change{ and the meaning for the key notations used in Sec.~\ref{sec:formalism} are summarized in Table \ref{tab:notation_summary}.  }

\begin{table*}[h!]
%  \resizebox{0.5\textwidth}{!}{\begin{minipage}{\textwidth}
  \caption{ Summary of the meaning of the  key notations  used in Sec.~\ref{sec:formalism}.  }
\label{tab:notation_summary}
\resizebox{\textwidth}{!}{\begin{minipage}{1.17\textwidth}
\begin{tabular}{ l| l }
\hline
\hline
Notation &  Meaning   \\
\hline
Latin index, e.g.~$i$                     &  The index for spec-$z$ sample bin $i$     \\ 
Latin index with a prime, e.g.~$i'$       &  The index for photo-$z$ sample bin $i'$      \\ 
%Greek index, e.g.~$\mu$                   &  The index for angular bin $\mu$            \\
$M_{i'}$ ($M_{i}$)                           &  The angular galaxy number density of the photo-$z$ sample in photo-$z$ bin $i'$ (spec-$z$ bin $i$)   \\
$N_{i}$                                   &  The angular galaxy number density of the spec-$z$ sample in spec-$z$ bin $i$  \\
$\bar{M}_{i'}$ ($\bar{M}_i$)               & The mean galaxy number density of the photo-$z$ sample in  photo-$z$ bin $i'$ (spec-$z$ bin $i$)   \\
$\bar{N}_{i}$                             & The mean galaxy number density of the spec-$z$ sample in spec-$z$ bin $i$   \\
$\delta_{i'} $ ($\delta_{i}$)                 &  The angular density fluctuation of the photo-$z$ sample in photo-$z$ bin $i'$ (spec-$z$ bin $i$)   \\
$\epsilon_{i} $                           &  The angular density fluctuation of the spec-$z$ sample in spec-$z$ bin $i$   \\

$Q_{ki'}$                                  &  The probability that a galaxy in spec-$z$ bin $k$ leaks to photo-$z$ bin $i'$   \\
$P_{ki'} $                                 & The fraction of galaxies in photo-$z$ bin $i'$ originating from spec-$z$ bin $k$    \\

$C_{i'j'}$                                 & The correlation function of the photo-$z$ sample between bin $i'$ and $j'$  \\
$C_{ij'}$                                 & The cross correlation function between the spec-$z$ sample in bin $i$ with the photo-$z$ sample in bin $j'$  \\
$C_{kk}'$                                 & The correlation function of the photo-$z$ sample in the spec-$z$ bin $k$    \\
$C_{kk}^{\rm x} $                           & The cross correlation function between the spec-$z$ sample in bin $k$ with the photo-$z$ sample in spec-$z$ bin $k$ \\
$C_{kk}^{\rm m} $                           & The matter  correlation function in spec-$z$ bin $k$   \\

$b_k' $                                   & The galaxy bias parameter of the photo-$z$ sample in spec-$z$ bin $k$ (no photo-$z$ bin dependence)  \\
$b_{k i'} $                               & The galaxy bias parameter of the photo-$z$ sample from photo-$z$ bin $i'$  in spec-$z$ bin $k$   \\
$b_k $                                   & The galaxy bias parameter of the spec-$z$ sample in spec-$z$ bin $k$   \\

$R_{ki'} $                                 & $b_k' P_{ki'} $, the quantity directly constrained by CZ  \\
%$\mathcal{J}_1  $ ($\mathcal{J}_2$)      & The SC (CZ) cosmost function \\

\hline \hline
\end{tabular}
\end{minipage}}
\end{table*}

Because of photo-$z$ uncertainties, galaxies within a photo-$z$ bin may originate from multiple spec-$z$ bins:
\beq
\label{eq:Mjp_Qijp_relation}
 M_{i'} =  \sum_k  Q_{k i'} M_k,
\eeq 
where $M$ is the angular galaxy number density,  $Q_{ki'}$ represents the probability that a galaxy in spec-$z$ bin $k$ leaks into photo-$z$ bin $i'$. The index with (without) a prime denotes the photo-$z$ (spec-$z$) bin index.

The correlation function of the photo-$z$ sample can be written as
\begin{align}
\langle M_{i'} M_{j'}  \rangle = \sum_{k,l}  \langle Q_{k i'} M_k Q_{l j'} M_l \rangle . 
\end{align}
By expressing $M_{i'} = \bar{M}_{i'}(1 + \delta_{i'})$ and $M_{i} = \bar{M}_i(1 + \delta_i)$ in terms of their respective means $\bar{M}$ and fluctuations $\delta$, we can write the angular overdensity correlation function as
\begin{align}
\label{eq:C_photoz}
C_{i'j'} \equiv \langle \delta_{i'}  \delta_{j'} \rangle & = \sum_k P_{ki'} P_{kj'} C'_{kk} , 
\end{align}
where $ C'_{kk} = \langle \delta_k \delta_k \rangle  $ is the spec-$z$ angular overdensity correlation function of the photo-$z$ sample in the spec-$z$ bin $k$.  We have defined  $ P_{ki'} $ as
\beq
\label{eq:Pijp_scatteringrate} 
P_{ki'} = \frac{  Q_{ki'} \bar{M}_k}{ \bar{M}_{i'} }. 
\eeq
From Eq.~\eqref{eq:Pijp_scatteringrate}, it is evident that $P_{k i'} $ satisfies the normalization condition
\beq
\label{eq:P_normalization}
\sum_k P_{ki'} = 1 .
\eeq
\change{ Thus $P_{ki'} $  represents the fraction of galaxies in photo-$z$ bin $i'$  coming from spec-$z$ bin $k$.   }

\change{By expressing $M_{i'}$ and $M_i$ in Eq.~\eqref{eq:Mjp_Qijp_relation} in terms of their fluctuations and then invoking Eq.~\eqref{eq:P_normalization},  we get the following relation }
\beq
\label{eq:deltajp_Pijp_relation}
\delta_{i'} = \sum_j P_{ji'} \delta_j,  
\eeq
which is the analogous relation to Eq.~\eqref{eq:Mjp_Qijp_relation}.  In fact, \change{$P_{ji'} $} is the true-$z$ distribution of the photo-$z$ sample, which is often denoted as $n(z)$. In this paper, we use these two notations interchangeably.

In arriving at Eq.~\eqref{eq:C_photoz},  we have assumed that $C_{ij}'$ is diagonal. In other words, the correlation between spec-$z$ bins is non-zero only when they are within the same bin. \change{ This is an excellent approximation and it is exact under  Limber approximation [e.g.~\citet{Simon2007}]}.  Eq.~\eqref{eq:C_photoz} is the essence of the SC method, and it enables us to extract  $P_{ki'} $ using the clustering information of the photo-$z$ data alone \citep{Schneider_etal2006, Zhang_etal2010}.

Assuming linear galaxy bias,  we can further write Eq.~\eqref{eq:C_photoz} in terms of the underlying spec-$z$ matter power spectrum  $C_{kk}^{\rm m} $ as 
\begin{align}
\label{eq:C_photoz_biasform}
C_{i'j'}  = \sum_k P_{ki'} P_{kj'}  b_{k}^{\prime 2}  C^{\rm m}_{kk}, 
\end{align}
where $b_{k}^\prime  $ is the bias parameter of the photo-$z$ sample in  $k$th spec-$z$ bin.

From Eqs.~\eqref{eq:Mjp_Qijp_relation}  to \eqref{eq:C_photoz}, we implicitly assume that the leakage due to photo-$z$ is universal irrespective of the galaxy type composition  in the spec-$z$ bin. To illustrate this point, let us assume that there are two types of galaxies in a spec-$z$ bin, say blue and red galaxies. The universal leakage assumption asserts that both types of galaxies leak to different photo-$z$ bins in the same proportion. Given photo-$z$ estimation is based on the photometry information, which is closely related to the galaxy types, this assumption can only hold approximately.

\change{ For the present work, the violation of the universal leakage assumption manifests as the dependence of the bias parameter on the photo-$z$ bin index $i'$.  }  Without the universal leakage assumption, Eq.~\eqref{eq:C_photoz_biasform} would be generalized to  
\begin{align}
\label{eq:C_photoz_biasform_nouniversality}
C_{i'j'}  = \sum_k P_{ki'} P_{kj'}  b_{k i'}  b_{k j'}  C^{\rm m}_{kk}, 
\end{align}
where $b_{k i^\prime }  $ is the bias parameter of the $i'$th photo-$z$ sample in the $k$th spec-$z$ bin. In this work, we assume that the leakage is universal.

%%%%%%%%%%%%%%%%%%%%%%%%%%%%%%%%%%%%%%%%%%%%%%%%%%%%%%%%%%%%%%%%%%%%%%%%%%%%%%%%%%%%%%%%

Suppose now we have another spec-$z$ sample with number density in spec-$z$ bin $i$, $N_i$, which  is related to its mean number density $\bar{N}_i $ and density fluctuation $\epsilon_i  $ by the relation $  N_i = \bar{N}_i ( 1 + \epsilon_i )  $.  We note that   $N_i$ and $ \epsilon_i $ are in general different  from the corresponding quantities for the photo-$z$ sample in spec-$z$ bin, $M_i$ and $\delta_i$. The spec-$z$ sample is often much less abundant than the photo-$z$ sample, and it tends to be brighter, and hence the linear bias of the spec-$z$ sample is likely to be higher.

The cross correlation function between the spec-$z$ and photo-$z$ sample reads
\begin{align}
\langle N_i M_{j'}  \rangle = \sum_{k}  Q_{k j'} \langle N_i M_k \rangle . 
\end{align}
Similar to the derivation of the SC results, we have the corresponding CZ equation:
\begin{align}
\label{eq:C_photoz_specz}
C_{ij'} \equiv \langle \epsilon_{i} \delta_{j'}  \rangle & = P_{ij'}  C_{ii}^{\rm x} \\
& = P_{ij'} b_i b_{i}^\prime   C_{ii}^{\rm m},
\end{align}
where $ C_{ii}^{\rm x} = \langle \epsilon_i \delta_i \rangle  $ is the cross angular correlation function \change{between the spec-$z$ sample in bin $i$ with the photo-$z$ sample in spec-$z$ bin $i$,}  and  $b_i$ is the bias parameter of the spec-$z$ sample in $i$th spec-$z$ bin.  %We stress that the correlation function of the photo-$z$ sample in spec-$z$ bin $C'_{ii}$ is different from the spec-$z$ correlation function because of different bias parameters. 
For completeness, without the universal leakage assumption,  we would instead have
\begin{align}
\label{eq:C_photoz_specz_nouniversal}
C_{ij'}  =  P_{ij'} b_i b_{ij'}  C_{ii}^{\rm m}.
\end{align}

Unlike the SC case, which is quadratic in $P$, CZ is a linear problem. In the usual CZ method, the bias parameter of the spec-$z$ sample, $b_j $ can be measured easily but it is difficult to get $b_{j}'$. Thus CZ only directly constrains $ P_{ij'} b_{i}' $ if we assume some theoretical matter correlation function model. Moreover, since $ P_{ij'}$ is normalized, if $b_{i}' $ is a constant, it has no effect on the estimation of $ P_{ij'}$.  However, if $b_{i}'$ evolves with redshift (or index $i$), its evolution is degenerate with $ P_{ij'}$.  Indirect methods have been proposed to mitigate the impact of bias evolution \citep{Schmidt_etal2013,Davis_etal2018,vandenBusch_etal2020, Cawthon_etal2022,Gatti_etal2022}.
From Eq.~\eqref{eq:C_photoz_biasform}, it is clear that the bias evolution issue also affects the SC problem.  When we allow  the correlation function in the spec-$z$ bin to be a free parameter (called $P$ method below), the degree of freedom due to  bias evolution is taken into account.

We shall test different approaches to solve the system of SC and CZ equations.  First we regard $P_{ij'}$, $C_{ii}^{\rm x}$, and $C'_{ii} $ as unknowns, denoted as the $ P $ method.  This is the most general parameterization, but the constraint is slightly weakened. Similar approach is taken in SC, e.g.~\cite{Zhang_etal2017}.   On the other hand,  \change{ analogous to the approach in CZ, } we  take $ R_{ij'} \equiv P_{ij'} b_{i}' $  as the variable \change{ and abbreviate it  as the $R$ method}. We note that there is no  normalization constraint on  $ R_{ij'} $.   If we assume that $b_i$  is known from measurement and $C_{kk}^{\rm m } $ can be computed theoretically, then we only need to solve for $ R $ in the system: 
\begin{align}
\label{eq:CZ_eq_Rformat}
C_{ij'} & = R_{ij'} b_i C_{ii}^{\rm m} ,\\
\label{eq:SC_eq_Rformat}
C_{i'j'} & = \sum_k R_{ki'} R_{kj'}  C_{kk}^{\rm m} . 
\end{align}
As a side note, in terms of the variable  $ R_{ij'} \equiv P_{ij'} b_{ij'} $, it is easy to see that the non-universality issue does not introduce extra complications in the case of CZ.

\subsection{ Solution to the SC and CZ equations}
\label{sec:numerical_sol}

\subsubsection{Cost function and multiplicative update rules} 

Our goal is to develop an efficient method to solve Eq.~\eqref{eq:C_photoz} and \eqref{eq:C_photoz_specz} simultaneously for $P_{ij'}$, $C'_{ii} $, and $C_{ii}^{\rm x} $.    Here we use the \change{$P$ method} as an example and  comment on the \change{ $R$  method} later on.  In particular, because Eq.~\eqref{eq:C_photoz} is a system of coupled quadratic equations, it is challenging to solve. Several studies have been devoted to solving the SC equation \citep{Erben_etal2009, Benjamine_etal2010, Benjamine_etal2013, Zhang_etal2017}.  Unlike prior \change{researches}, \citet{Zhang_etal2017} solves Eq.~\eqref{eq:C_photoz} in full generality, without limiting the analysis to a two-bin scenario  \citep{Erben_etal2009}  or relying on a linear coupling  (in $P$) approximation as done by \cite{Benjamine_etal2010}. In this work, inspired by \citet{Zhang_etal2017}, we construct multiplicative update rules to solve  Eq.~\eqref{eq:C_photoz} and \eqref{eq:C_photoz_specz} jointly.

We aim to derive an iterative update rule to minimize the sum of the SC and CZ cost functions: \footnote{This is actually the $\chi^2$ with the regularization by the weight function $W$, but here we follow the previous convention \citep{LeeSeung_2000,Zhang_etal2017} to call it $ \mathcal{J} $. } 
\begin{equation}
  \label{eq:J1_J2_cost}
    \mathcal{J}=\mathcal{J}_{1}+\mathcal{J}_{2},
\end{equation}
where $\mathcal{J}_{1}$ and $\mathcal{J}_{2}$ are the contributions from SC and CZ respectively.  They are given by 
\begin{align}
  \label{eq:J1_cost}  
    \mathcal{J}_{1} & =\frac{1}{2} \sum_{i^{'},j^{'}, \mu }  \frac{ W_1(\theta_\mu) }{ \sigma_{i'j'}^2 (\theta_\mu) }  \Big[ D_{i^{'}j^{'}}(\theta_\mu)-\sum_{k}P_{ki^{'}}P_{kj^{'}}C'_{kk}(\theta_\mu)  \Big]^{2} ,\\
  \label{eq:J2_cost}  
    \mathcal{J}_{2} & =\frac{1}{2} \sum_{i,j^{'}, \mu}  \frac{ W_2(\theta_\mu) }{ \sigma_{ij'}^2 (\theta_\mu) }  \Big[ D_{ij^{'}}(\theta_\mu )-P_{ij^{'}}C_{ii}^{\rm x}(\theta_\mu)  \Big]^{2} ,  
\end{align}
where $D$ denotes the data measurement (angular correlation function in our case), $\sigma$ is the error bar of the measurement, and $W_1$  and $W_2$  are the weight functions. We consider weight function of the form $\theta^{n}$ in this work [see also \cite{Menard_etal2013}].   Here we use Latin indices for the redshift bins and Greek indices for the angular bins.  The update rule will update $ P_{ij^{'}} $, $C_{ii}' $, and $C_{ii}^{\rm x} $ iteratively to look for a minimum of $\mathcal{J}$.  Additionally, we will consider assigning different weights to $\mathcal{J}_1 $ and $\mathcal{J}_2 $ later on.

By minimizing the cost function with respect to $P_{ab'} $, a multiplicative update rule [Eq.~\eqref{eq:update_rule_Pijp}] for $P_{ab'} $ can be derived. Upon multiplying the factor in Eq.~\eqref{eq:update_rule_Pijp} to   $P_{ab'} $ repeatedly,   $P_{ab'} $ converges to the solution. This method can be viewed as a variant of the gradient decent \citep{LeeSeung_2000}.   Similar update rule can also be established for $ C_{ii}^{\rm x}$ and $C'_{ii} $ [Eq.~\eqref{eq:update_rule_Cii}]. The details of the derivation are relegated to Appendix \ref{Appendix:Derivation_update_rule}.  A few comments of the multiplicative rules are in order.

First,~\cite{Zhang_etal2017} employed the non-negative matrix factorization (NMF) algorithm, originally proposed by~\cite{LeeSeung_2000}, to address the SC problem. This approach has been adopted in subsequent studies~\citep{Peng_etal2022, XU2023, Song_etal2024, PengYu_2024}.   Due to the non-negativity constraint inherent in the SC model, the NMF method offers an elegant and efficient solution. %, achieving a redshift estimation accuracy in the range of 0.01\%--1\%.
NMF assumes that the model can be factorized into a product form $W H$, where $W$ and $H$ are distinct matrices. To apply this framework to the SC model,~\cite{Zhang_etal2017} reformulated it as $W H_\theta$, with $W=P^T$ and $H_\theta=C(\theta) P$.  However, this splitting is somewhat artificial for the SC problem and may cause difficulty in including the CZ information  due to its rigid structure.
%However, as the matrix form of the model becomes more complex, the splitting method may no longer be applicable, potentially leading to numerical instability.
To further optimize and generalize the approach for solving the SC model, we bypass the NMF interpretation and instead treat it directly as a minimization problem with respect to $P_{ab'} $. By employing a specialized parameter update scheme inspired by the NMF approach, our simplified interpretation allows for an efficient combination of information from both SC and CZ models.

%First, \cite{Zhang_etal2017} adopted the non-negative matrix factorization (NMF) algorithm from \citet{LeeSeung_2000} to solve the SC problem. This route has been followed in subsequent studies \citep{Peng_etal2022, XU2023, Song_etal2024, PengYu_2024}. The primary attractive feature of the NMF approach is its power in dimension reduction, i.e.~approximating the full model by a model with  smaller number of degrees of freedom.   But this feature does not seem to be particularly beneficial to the SC problem since we have a well defined ``fixed'' model in this case .  The NMF assumes that the model can be factorized into the product form $WH$ with $M$ and $H$ being distinct matrices.  In order to cast the SC model in this form, \cite{Zhang_etal2017} writes the SC model as  $W H_\theta $ with $ W =  P^T $ and $H_\theta = C(\theta)  P $. This split is somewhat artificial and may cause difficulty in generalization due to its rigidity.  Instead, we bypass the NMF interpretation and treat it as a minimization problem with respect to~to $P_{ab'} $ directly. Our straightforward interpretation enables us to combine the information of SC and CZ easily. 

Secondly, the cost function  for the multiplicative update rule is often taken to be the mean squared error without the inverse error bar weighting, e.g.~\citet{LeeSeung_2000, Zhang_etal2017}. Treating all the data on equal footing is at best sub-optimal.  We have to include the  error bar to down-weight the contribution of the poor measurements and up-weight the good ones,  and to eliminate the impact of missing data measurements.  Here we have improved over previous treatments by taking into account the error of measurements and the inclusion of additional weighting functions.  The proper treatment should include the full covariance matrix. However, in the derivation of the multiplicative rule, we have to separate the positive part of the gradient from the negative one. Usage of the full covariance will hinder this separation. Thus for simplicity, we opt to use diagonal error bars in Eqs.~\eqref{eq:J1_cost} and \eqref{eq:J2_cost}.  \change{ Under the diagonal covariance approximation, the data points are taken to be independent. This may seem a poor approximation, but we observed that the overall performance of the algorithm is good.  }
We mention that \cite{XU2023} used $ \chi^2$ as the stopping criterion of the iteration process, but the update rule is still based on the cost function without the inverse error weighting. Recently, \cite{PengYu_2024} independently proposed to include error weighting in the NMF algorithm to solve the SC problem. They adopted the improved NMF algorithm  of  \cite{Zhu2016, GreenBailey_2023}, which generalizes the \cite{LeeSeung_2000} NMF update rule to account  for the data measurement error.  We expect its performance to be similar to our SC results.

Thirdly, the multiplicative update rule ensures that the estimated values are non-negative provided that the measurements are non-negative \footnote{ We mention that the suppressed Gaussian process can also be tuned to return non-negative estimated values \citep{Naidoo_etal2023}. } . This is less likely to hold for $C_{ij'}$ (see Fig.~\ref{fig:clustering_measurements} below) due to its larger covariance. Negative measurements may spoil the update rule and result in negative estimated value. To mitigate this problem,  in \cite{XU2023}, negative measurements are set to small positive values by hand, and a more refined process is applied in  \cite{PengYu_2024}.   %Anyway, by applying the inverse error weighting, the negative measurements due to large fluctuations are suppressed to some extent.
In Appendix \ref{sec:Impact_NoNegative}, we test the impact of setting the negative measurements to a tiny positive value.

Finally, we mention that our update rule is also applicable when $\mathcal{J}_{1} $ or $ \mathcal{J}_{2} $ are missing. In particular it can be used for the SC method. We will contrast ours against the results from \cite{XU2023} below.

Before closing this section, we comment on the solution of the system Eqs.~\eqref{eq:CZ_eq_Rformat} and \eqref{eq:SC_eq_Rformat}  \change{ in the $R$ method}. Basically it is the same as the \change{ $P$ method}, except that $R_{ij'}$ is solved without the normalization constraint, viz.~Eq.~\eqref{eq:Pabp_update_nonorm}. From  $R_{ij'}$, if we assume that there is no bias evolution, then using Eq.~\eqref{eq:P_normalization}, we have \footnote{ \change{ In Eq.~\eqref{eq:bjp_noevolution} the indices on both sides do not match. It will look more pleasant if we employ the general bias form  $ b_{ij'} $. }  } 
 \begin{align}
 \label{eq:bjp_noevolution}
 b_{i}' & \approx b' = \sum_i R_{ij'},
 \end{align}
 and it follows that 
 \begin{align}
\label{eq:Pijp_noevolution}
P_{ij'} & =  \frac{ R_{ij'} }{  \sum_k R_{kj'} }. 
 \end{align}
%A consequence of the no bias evolution approximation is that $b'$ is independent of $i$ or $j'$. We shall test this approximation below.   

\begin{figure*}[htb!] 
    \includegraphics[width=\linewidth]{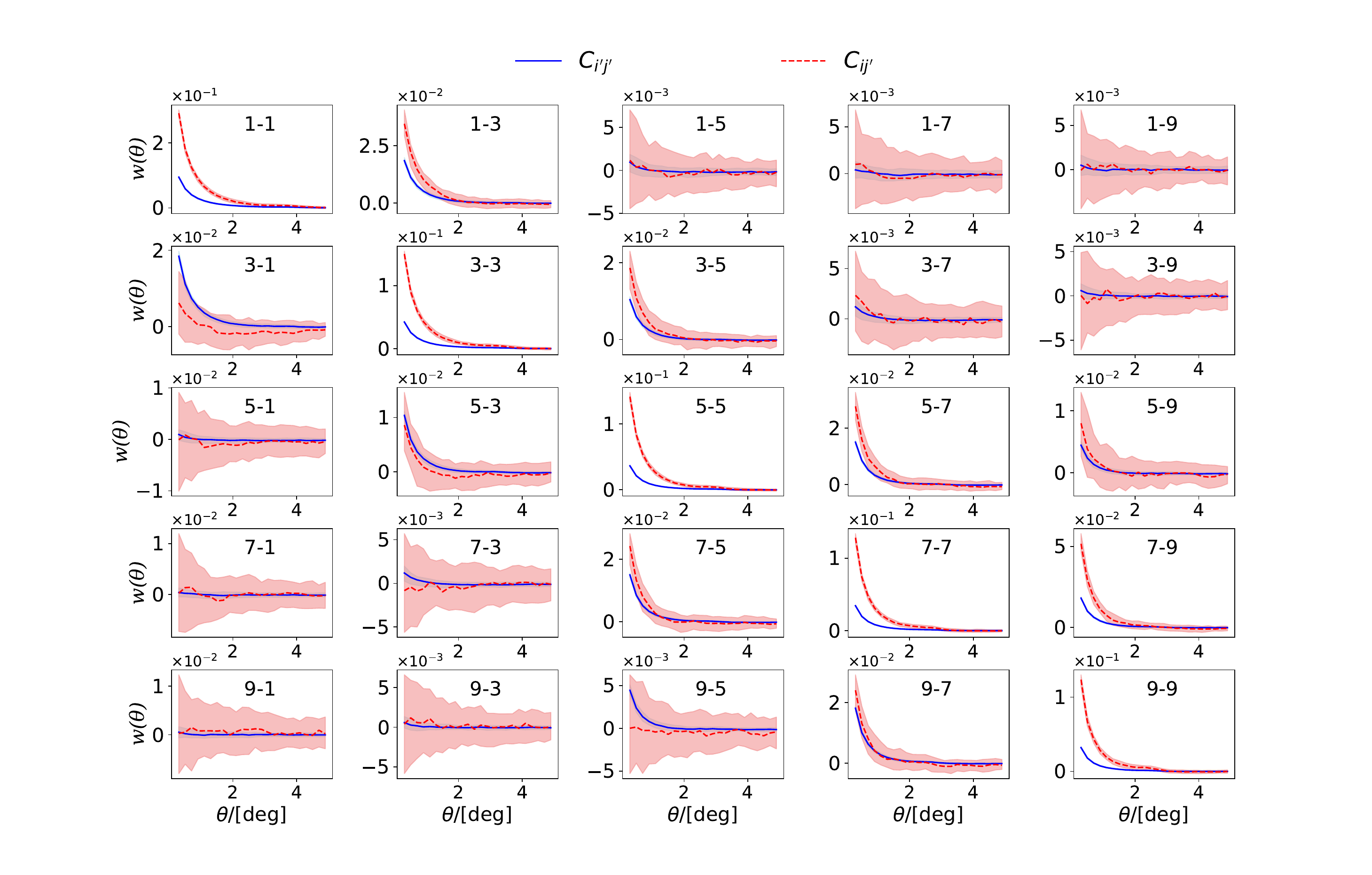}
    \caption{  Sample of the photo-$z$ angular correlation function between the photo-$z$ bin $i'$ and $j'$, $ C_{i' j'} $ (blue) and  the cross angular correlation function between the spec-$z$ bin $i$ and photo-$z$ bin $j'$,  $ C_{i j'} $ (red) to be used for true-$z$ inference.   The line and its associated color band are the median and 16 and 84 percentile among 100 mock runs. The whole redshift range [0.6,1.1] is divided into 10  photo-$z$ bins, each of width  $\Delta z_{\rm p}  = 0.05 $. This  results in a true-$z$ distribution with resolution of $ \Delta z = 0.05 $. The label $i$-$j$ represents $i'j'$ for  photo-$z$ correlation function and $i j'$ for spec-$z$-photo-$z$ cross correlation function respectively. We only show the odd bin results for clarity.  } 
    \label{fig:clustering_measurements}
\end{figure*}

\subsubsection{Numerical implementation}

%\subsubsection{ Review of the previous SC algorithm }

%\KCC{Which parts are due to \citet{Zhang_etal2017}, which one is due to \citet{XU2023} or \cite{Peng_etal2022}? }

Because our method is rooted in the previous works and  we shall contrast the old result with ours, it is helpful to review the algorithm of \citet{Zhang_etal2017} and its improvements in \cite{XU2023}. We shall denote this as the ``Old NMF'' and use it as a benchmark for comparison.

The algorithm starts by initializing $P_{ij'} $  randomly as a diagonally dominant matrix.  The initialization procedure assumes that the true-$z$ distribution peaks at the photo-$z$ estimate and decreases monotonically on both sides.  The initial $C'_{ii} $ and  $C_{ii}^{\rm x} $ are obtained from Eqs.~\eqref{eq:C_photoz} or \eqref{eq:C_photoz_specz} using the initial $P_{ij'}$.  Before applying the multiplicative update rule, \cite{Zhang_etal2017} found that it was necessary to get a preliminary solution using a fixed-point method for the NMF step to be successful. Using this preliminary solution as the initial trial solution,  the NMF update rule is applied until a minimum of the cost function is attained.  \cite{Zhang_etal2017} writes the SC model as  $W H_\theta $ with $ W =  P^T $ and $H_\theta = C(\theta)  P $.   In each step, the update rule analogous to that in  \cite{LeeSeung_2000} is applied to $W$ with  $H_\theta $ fixed. The $P$ in $ H_\theta $ is then replaced  with the new  $W$, and with $P$ held fixed $C(\theta) $ is subsequently solved by the least square solution.  To estimate the error bar  on $P_{ij'}$, \cite{XU2023} generates 100 sets of $P_{ij'} $ and for each set, the measurement $C_{i'j'} $ is perturbed by adding a Gaussian perturbation obtained by sampling the covariance of  $C_{i'j'}$.   We call this set random runs to distinguish them from the mock realizations.  The best fit and the $1\sigma $ error are estimated by  the median and the half width between the 16 and 84 percentile, respectively.

We shall contrast three different setups: SC, CZ, and the joint inference by SC and CZ, denoted as SC+CZ. For SC and SC+CZ, the default method uses $P_{ij'}$ and $C_{ii}'$ (and $C_{ii}^{\rm x} $) as unknowns (abbreviated as $P$ method), while for CZ, we use $R_{ij'} $ as variables (denoted as $R$ method). The reason for these differences is related to the error estimation and we shall comment on it later on.  For the $P$ method, we initialize $P_{ij'} $ and $ C'_{ii} $ (and  $C_{ii}^{\rm x}$)  as in the Old NMF, but we then feed the initial guess to the multiplicative update rule  [Eqs.~\eqref{eq:update_rule_Pijp} and \eqref{eq:update_rule_Cii}] directly because we find that our algorithm no longer requires the intermediate solution from the fixed point method.  We take the weighting function to be $\theta^{n} $ with $n=-1$.   Similar to \cite{XU2023}, we generate 100 random runs to estimate the best fit and $1\sigma$ error bar.  The best fit and $1\sigma$ error bar are estimated by the median and half width of the 68-percentile about the median. In each run, we also perturb the measurement using a Gaussian perturbation derived from the mock covariance.

For the $R$ method, the procedures are similar. To initialize  $R_{ij'}$ we generate the initial  $P_{ij'}$ as above, but we find that the result is sensitive to the trial $b_i'$. Thus we perform a grid search on $ b_i' $ to find the one that gives rise to the minimum $ \mathcal{J} $ and use it as the final solution.  We then get  $P_{ij'}$ from the best fit $R_{ij'}$  using Eq.~\eqref{eq:Pijp_noevolution}.

\section{ Mock test results }
\label{sec:mock_test}

In this section, we apply our algorithm to the mock catalog to test its performance.

\subsection{ Description of the mock catalog}

We shall test and validate our results using mock catalogs.  For this purpose, we employ the ICE-COLA mocks \citep{DES:2021fie}, which is  a specialized mock catalog tailored for the DES Y3 BAO analyses \citep{ DES:2021esc}. A brief overview is provided here, with further details available in \citet{DES:2021fie}.

The ICE-COLA mocks are derived from the COLA simulations, built upon the COLA method \citep{Tassev_etal_2013}, and executed through the ICE-COLA code \citep{ice-cola}. The COLA method combines the second-order Lagrangian perturbation theory with the particle-mesh simulation technique, ensuring the preservation of accuracy in large-scale modes despite the utilization of coarse simulation time steps.  In each simulation, there are  $2048^3$ particles in a cube measuring side length of 1536 $\MpcOh$. The comoving simulation is transformed to a lightcone simulation extending up to $z \sim 1.4 $.  % The mass resolution is aligned with the MICE Grand Challenge $N$-body simulations \citep{Fosalba_etal2015, Crocce_etal2015}.
The cosmology adopted by the mock catalog follows that in the MICE simulation \citep{Fosalba_etal2015, Crocce_etal2015},  which is a flat $ \Lambda$CDM with $\Omega_{\rm m} = 0.25$,  $\Omega_{\Lambda} = 0.75$, $h=0.7$, and $\sigma_8 = 0.8$.  Each mock occupies a DES Y3 footprint, covering 4180 deg$^2$ in area.  We make use of 100 mock catalogs to estimate the covariance. \change{ The same set of mocks are also used to estimate the ensemble mean and its associated  error. }

The mock galaxies are allocated to the ICE-COLA halos through a hybrid method combining Halo Occupation Distribution and Halo Abundance Matching, mirroring the technique outlined in \citet{Avila:2017nyy}. %These mocks are fine-tuned by aligning them with the redshift distribution and angular clustering amplitude of the actual galaxy samples.
These mock galaxies resemble the red galaxy sample in DES Y3 (see \cite{Carnero_etal2012} for more details on this sample).  The mock galaxies are equipped with realistic photo-$z$. To do so, a 2D joint probability distribution in the photo-$z$ and spec-$z$ space is constructed using a sample of actual galaxies possessing both types of redshifts.  Through sampling this distribution, the candidate mock galaxies are assigned the appropriate photo-$z$'s.

Because the galaxies have both photo-$z$ and spec-$z$ labels, we can use them to create photo-$z$ and spec-$z$ samples.  As in DES Y3, we select galaxies in the photo-$z$ range [0.6,1.1] and divide them into five tomographic bins, each of width 0.1. Our goal is to determine the true-$z$ distribution of the samples in these tomographic bins using clustering-based methods.  To model the fact that the spec-$z$ galaxies are generally few and bright,  we approximate it by selecting the most massive galaxies from the mock. In this case we expect  the clustering amplitude of the spec-$z$ sample to be  higher than the intrinsic clustering of the photo-$z$ sample since the linear bias increases with mass. We consider samples containing the most massive $p$\% of the galaxies with $p=5$, 1, and 0.5.

When we restrict the sample to the photo-$z$ range [0.6,1.1], a small fraction of the galaxies will possess spec-$z$ values outside this range. This implies that there are reductions in correlation signal from SC and CZ. This problem can be alleviated by considering the redshift range of data, \change{ both photo-$z$ and spec-$z$ data, to be   sufficiently wide relative to } the redshift range of interest. Because the mock data are only available in the photo-$z$ range  [0.6,1.1], this is not possible.   Here we clean the sample by removing the galaxies with spec-$z$ values outside of the range [0.6,1.1], and this effectively forces the true-$z$ distribution to vanish outside the range [0.6,1.1].  This cleaning cannot be done in real data. In Appendix \ref{sec:Impact_OutlyingGal}, we study the impact of the outlying galaxies on the true-$z$ estimation.

\begin{figure*}
    \centering
    \begin{minipage}{0.48\linewidth}
   \includegraphics[width=\linewidth]{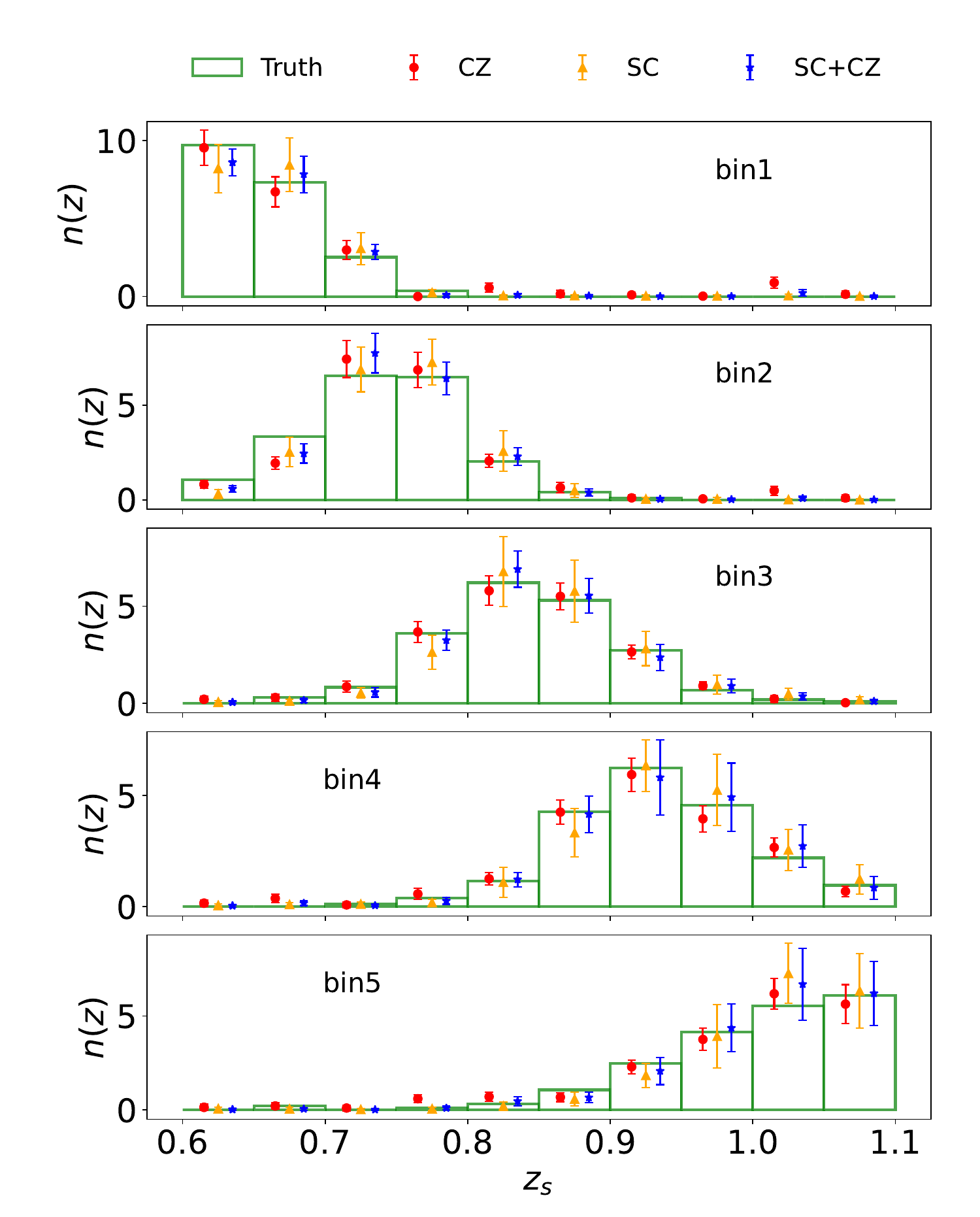}
    \caption{ True-$z$ distribution inferred by the clustering-based estimators [CZ (red), SC (orange), and SC+CZ (blue)]   are compared with the direct mock measurement (green bars). The clustering-based estimator data points are offset horizontally for clarity. The results for five tomographic bins are shown (from top to bottom). } % SC+CZ gives a  more accurate estimate of the true-$z$ distribution than CZ or SC alone.   }
    \label{fig:histogram_sc_cz_sc+cz_logM_1p}
    % \vspace{63pt}
    \end{minipage}
    \begin{minipage}{0.01\linewidth}
       \hspace{0.01\linewidth}
    \end{minipage}
    \begin{minipage}{0.48\linewidth}
        \includegraphics[width=1\linewidth]{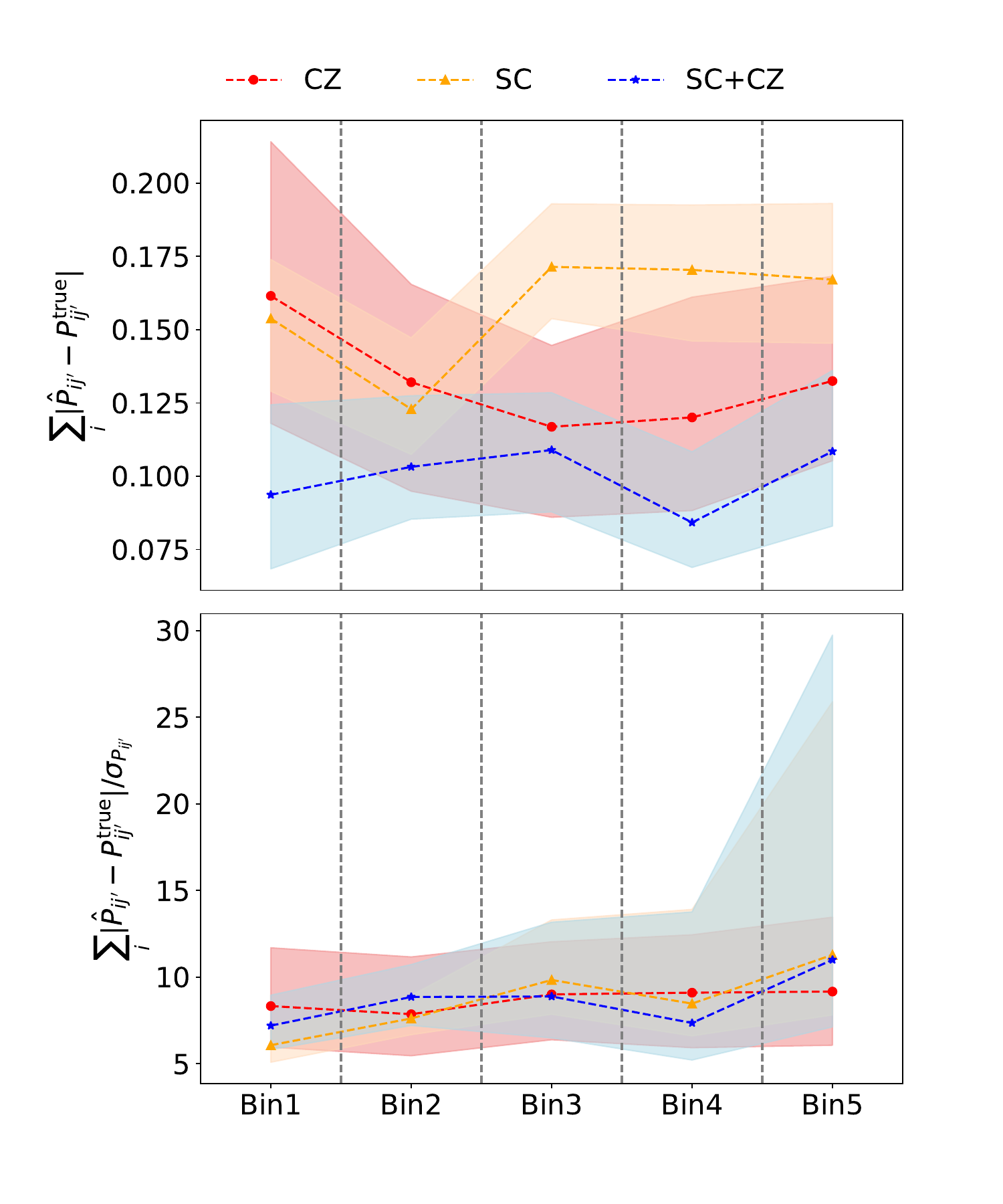}
        \caption{  Comparison of the accuracy of the true-$z$ distribution inferred using CZ (red), SC (orange), and SC+CZ (blue). {\it  Upper panel}:  $ \sum_i |\hat{P}_{ij'} - P^{\rm true}_{ij'} | $, the absolute difference between the true-$z$ distribution measured from the mock, $P^{\rm true}_{ij'} $  and the one estimated, $ \hat{P}_{ij'} $. {\it Lower panel}: $  \sum_i |\hat{P}_{ij'} - P^{\rm true}_{ij'} | / \sigma_{P_{ij'}}$ the absolute difference normalized by the estimated error  $\sigma_{P_{ij'}}$.  The line and color band represent the median and the 16 and 84 percentile among 100 mock runs.  The results for five tomographic bins are shown.  }
    \label{fig:two_piece_sc_cz_sc+cz_logM_1p}
    %\vspace{63pt}
    \end{minipage}
\end{figure*}

\begin{table*}[h!]
  \caption{  Bias in the mean redshift $e_z$ obtained with various methods.     }
\label{tab:mean_shift}
%\resizebox{0.7\textwidth}{!}{\begin{minipage}{\textwidth}
\begin{tabular}{l| lllll|}
\hline
\hline
Estimation method & \multicolumn{5}{c|}{  $ ( \bar{z}_{ \hat{n} }  -  \bar{z}_{n_{ \rm true }} )/(  1 +  \bar{z}_{n_{ \rm true }} )  \times 10^3  $ }   \\
                 &  1 & 2 & 3 & 4 & 5                       \\
\hline
SC                                   & 4.7 $\pm$ 1.2  & 5.8 $\pm$ 1.4  &  6.4 $\pm$ 1.4 &  6.3 $\pm$ 1.4   & 6.1 $\pm$ 1.8   \\ \hline
CZ  (0.5\%)                          & 21.3 $\pm$ 4.6  & 9.6 $\pm$ 3.3  & 0.5 $\pm$ 2.3  & -4.4 $\pm$ 2.6  & -7.5 $\pm$ 4.6  \\
CZ  (1\%)                            & 12.2 $\pm$ 5.9   & 4.9 $\pm$ 2.8  &  -0.5 $\pm$ 2.5  &  -2.1 $\pm$ 2.6   & -2.3 $\pm$ 4.1   \\
CZ  (5\%)                            & 9.0 $\pm$ 4.8  &  3.9 $\pm$  2.5   & -0.3 $\pm$ 1.8  & -1.5 $\pm$ 1.9  & -2.2 $\pm$ 2.7  \\ \hline
SC+CZ (0.5\%)                        & 3.6 $\pm$  2.0   & 3.4 $\pm$ 1.5  & 2.9 $\pm$ 1.4  & 3.1 $\pm$ 1.4  & 3.9 $\pm$ 2.0  \\
SC+CZ (1\%)                          & 3.6 $\pm$ 2.0   &  3.2 $\pm$ 1.5   & 2.4 $\pm$ 1.3 & 2.5 $\pm$ 1.3  & 3.5 $\pm$ 2.2    \\
SC+CZ (5\%)                          & 3.4 $\pm$  2.3  & 2.4 $\pm$  1.5  & 1.5 $\pm$ 1.2  & 1.5 $\pm$ 1.2   & 2.5 $\pm$ 1.7    \\ \hline
SC, Old NMF                          & 12.8 $\pm$ 4.7  & 17.1 $\pm$ 5.0  & 18.8 $\pm$  3.9   & 17.1 $\pm$ 4.9  & 11.8 $\pm$ 4.0  \\ \hline
SC+CZ (1\%), No spec-$z$ bin 9-10    & 3.3 $\pm$  1.7  & 3.0 $\pm$ 1.3  & 2.6 $\pm$ 1.3  &  2.5 $\pm$ 1.4  & 3.2 $\pm$ 2.3   \\
SC+CZ (1\%), No spec-$z$ bin 7-10    & 3.1 $\pm$  1.4  & 2.8 $\pm$ 1.4  & 2.5 $\pm$ 1.2  &  3.4 $\pm$  1.4 & 4.5 $\pm$ 2.1           \\
\hline \hline
\end{tabular}
\tablefoot{  The bias in the mean redshift $e_z$, which is given by the difference between the mean redshift computed with the true-$z$ distribution from various estimators  $\bar{z}_{ \hat{n} }$ and  the direct mock measurement $\bar{z}_{\rm true}$, and further divided by $1+\bar{z}_{\rm true}$   (and multiplied by $10^3$). We show the results for the 5 tomographic bins. The ensemble mean and the 1$\sigma$ are estimated from 100 mock runs. Although in some of the bins, CZ gives a smaller bias,  SC+CZ results are more stable and their total bias across the bins is smaller.  }
%\end{minipage}}
\end{table*}

\subsection{Clustering measurements}

The data for inference are the angular correlation functions. While the bin width of the target photo-$z$ sample is $ \Delta z_{\rm p} =0.1 $,  to increase the redshift resolution of the true-$z$ distribution, we  divided the sample into ten  photo-$z$ bins, each with bin width of $ \Delta z_{\rm p} =0.05 $.  From the true-$z$ inference code, we ended up with ten true-$z$ distributions with a redshift resolution of $\Delta z = 0.05 $ for these ten photo-$z$ samples.  We then combined every two true-$z$ distributions to get the weighted mean for our target photo-$z$ samples, e.g., the first and second true-$z$ distributions are combined to get the one for the first target photo-$z$ sample.

In Fig.~\ref{fig:clustering_measurements}, we show a sample of the angular correlation function  $w(\theta)$ to be used in true-$z$ inference. The plot shows  the photo-$z$ angular correlation function  $ C_{i' j'} $ and  the cross angular correlation between the spec-$z$ and photo-$z$ sample  $ C_{i j'} $.   In this plot, the spec-$z$ sample consists of the most massive 1\% galaxies. We have plotted the median and the $1\sigma$ band estimated by the 16 and 84 percentile among 100 mock runs. We only show the results for $i$-$j$ with $i$ and $j$ being odd for clarity.

These angular correlation function measurements are performed using CUTE \citep{Alonso_CUTE} with the grid method. We  compute them in the angular range of  $[0.2, 5]^\circ$  with linear binning of width   $ 0.2^\circ $.

%% \begin{figure}[htb!] 
%%       \includegraphics[width=\linewidth]{histogram_sc_cz_sc+cz_logM_1p.pdf}
%%     \caption{ The true-$z$ distribution inferred by the clustering estimators [CZ (red), SC (orange), and SC+CZ (blue)]   are compared with the direct mock measurement (green bars). The clustering estimator data points are offset horizontally for clarity. The results for five tomographic bins are shown (from top to bottom). } % SC+CZ gives a  more accurate estimate of the true-$z$ distribution than CZ or SC alone.   }
%%     \label{fig:histogram_sc_cz_sc+cz_logM_1p}
%% \end{figure} 

%% \begin{figure}[htb!] 
%%   %\includegraphics[width=1\linewidth]{5bins/two_piece/two_piece_sc_cz_sc+cz_logM_1p.pdf}
%%   \includegraphics[width=1\linewidth]{two_piece_sc_cz_sc+cz_logM_1p.pdf}
%%     \caption{  Comparison of the accuracy of the true-$z$ distribution inferred using CZ (red), SC (orange), and SC+CZ (blue). {\it  Upper panel}:  $ \sum_i |\hat{P}_{ij'} - P^{\rm true}_{ij'} | $, the absolute difference between the true-$z$ distribution measured from the mock, $P^{\rm true}_{ij'} $  and the one estimated, $ \hat{P}_{ij'} $. {\it Lower panel}: $  \sum_i |\hat{P}_{ij'} - P^{\rm true}_{ij'} | / \sigma_{P_{ij'}}$ the absolute difference normalized by the estimated error  $\sigma_{P_{ij'}}$.  The line and color band represent the median and the 16 and 84 percentile among 100 random runs.  The results for five tomographic bins are shown.  }
%%     \label{fig:two_piece_sc_cz_sc+cz_logM_1p}
%% \end{figure} 

\subsection{ Inference of the true-$z$ distribution }

We  show the results on the inference of the true-$z$ distribution in Fig.~\ref{fig:histogram_sc_cz_sc+cz_logM_1p}. We compare the results obtained with SC, CZ, and SC+CZ against the true-$z$ distribution measurement from the mock catalog. Here we show the results from a single mock.  They  are produced using the fiducial setup, and the spec-$z$ sample is the most massive 1\%.  % Visually, it seems that SC+CZ produces better results than SC or CZ alone. 

In order to quantify the accuracy of the results, \change{ we use all 100 mocks and } consider the metric  $ \sum_i |\hat{P}_{ij'} - P^{\rm true}_{ij'} | $, the absolute difference between the mock measurement $P^{\rm true}_{ij'} $  and the estimated result $ \hat{P}_{ij'} $.  We show this in the upper panel of Fig.~\ref{fig:two_piece_sc_cz_sc+cz_logM_1p}. The line corresponds to the median and the color band demarcates  the 16 and 84 percentile lines among 100 mock runs.

We find that  for the first two bins, SC and CZ results are similar with SC marginally better, and the accuracy of CZ  remains good for the high $z$ bins while SC deteriorates.  In reality, the number of spec-$z$ galaxies may plummet faster than the constant fraction selection assumed here, and thus the performance of CZ in the high $z$ bins may not be as good as the case shown here.  The combination SC+CZ achieves the minimal error, and performs better than SC or CZ alone. %This demonstrates  that in first two bins, the information is dominated by SC, while CZ plays the dominant role in the last three bins.  

To go on to check the accuracy of the error estimate on $ P_{ij'} $, $\sigma_{P_{ij'} }$, we plot the absolute difference normalized by the estimated error,  $  \sum_i |\hat{P}_{ij'} - P^{\rm true}_{ij'} | / \sigma_{P_{ij'}}$ in the lower panel of Fig.~\ref{fig:two_piece_sc_cz_sc+cz_logM_1p}. A rough estimate is that the difference should be of order $\sigma_{P_{ij'}} $, and so the sum is  $  \sim 5 $. We find that the result are generally larger than this simple estimate  by a factor of $\sim 1.5$, and it gets larger for the last bin. The large fluctuation of the 1$\sigma$ band especially for the last bin is caused by the occasionally too small error bar estimation in the tail of true-$z$ distribution.

\begin{figure*}[htb!] 
     \includegraphics[width=\linewidth]{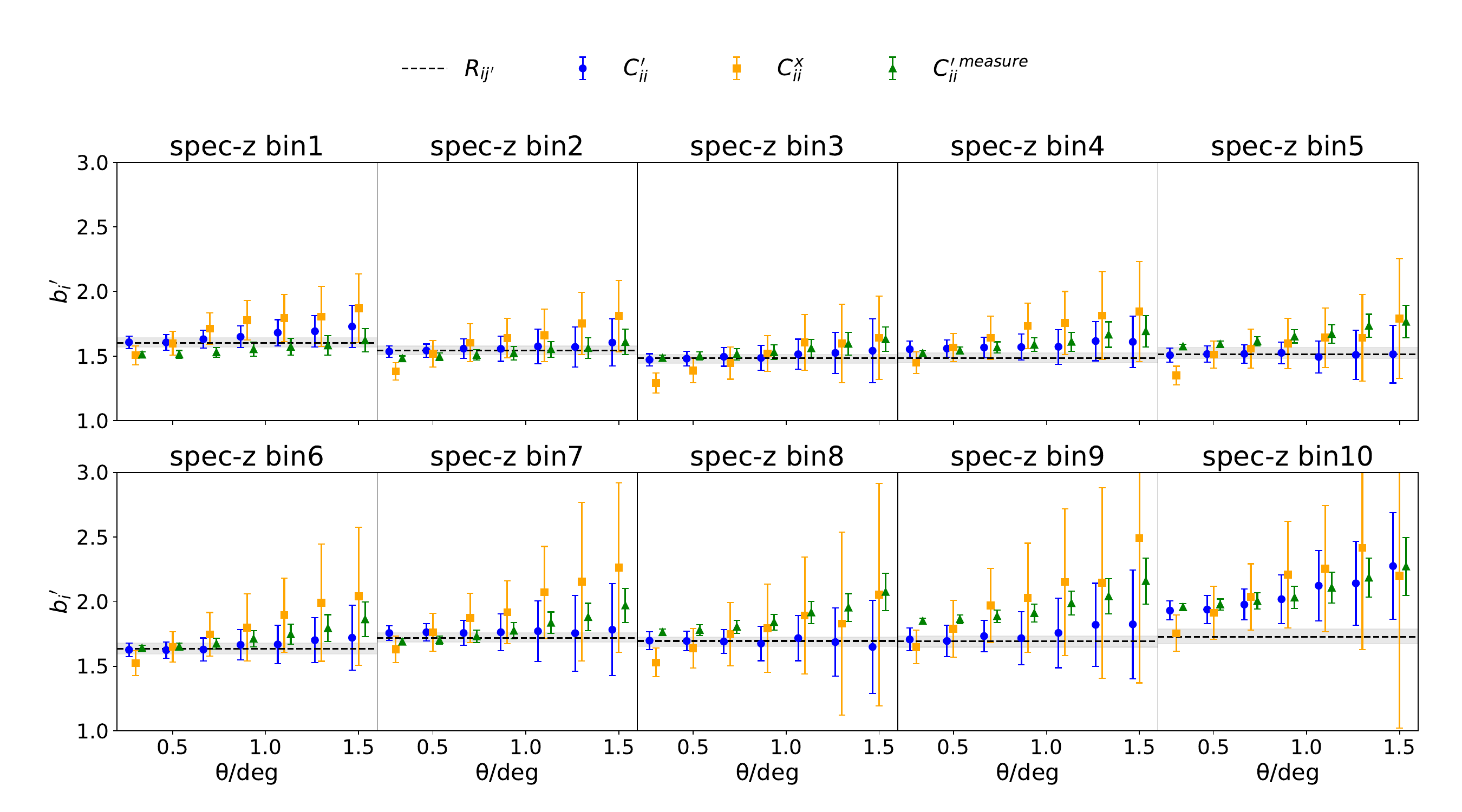}
     \caption{ Galaxy bias parameter of the photo-$z$ sample in the spec-$z$ bin $i$, $b_i'$ estimated by various methods.  The estimate by  $C_{ii}' $ (blue circles), $C_{ii'}^{\rm x}$ (orange squares), and $R_{ij'}$ (black dashed line with 1$\sigma $ error band in gray) are compared with the direct measurements (green triangles). The results for 10 spec-$z$ bins are shown. The data points are offset slightly horizontally for clarity.   See the text for details.  }
     \label{fig:bias}
\end{figure*}

\change{ To determine the true-$z$ distribution, all the cross bin clustering information is used; thus the information is not localized to a particular bin and it is hard to get a ``simple'' explanation of the trend seen in  Fig.~\ref{fig:two_piece_sc_cz_sc+cz_logM_1p}.  For CZ, all the spec-$z$ sample is used to cross correlate with a photo-$z$ sample.  Thus we argue that the photo-$z$ sample property could give a better estimate of the quality of the CZ inference.   After cleaning, the number of photo-$z$ galaxies in the photo-$z$ bins are 1228413, 1576759, 1683208, 1266897, and 708319 respectively. The effective bias parameters of the photo-$z$ sample in the tomographic bins are similar (about 1.1), but at both ends the bias is $\sim 1.3 $. The rise at both ends is caused by the cleaning process, which cuts off the true-$z$ distribution and hence increases the clustering amplitude [\citet{Chan_etal2024}]. Nonetheless the biases in the true-$z$ bins, $b_i' $ are quite constant (see Fig.~\ref{fig:bias}). Hence the number of photo-$z$ galaxies in the tomographic bins can qualitatively explain the trend of CZ in Fig.~\ref{fig:two_piece_sc_cz_sc+cz_logM_1p}.   The trend of the SC is even harder to interpret as it relies on the auto and cross clustering information all the bins.   }

%, but SC and SC+CZ are generally larger except for the first bin. Their $1\sigma$ bands are also large, especially for the third and fifth bins.  By inspecting Fig.~\ref{fig:histogram_sc_cz_sc+cz_logM_1p}, we deduce that the problem is caused by the occasionally very small error bar estimate  in the tail of the $n(z)$ distribution. \KCC{ Comment on other estimation of error bars using other methods } 

The bias in the mean of the true-$z$ distribution is commonly used as an indicator for the accuracy of the characterization of the true-$z$ distribution.\footnote{ For example, the target bias level of Euclid is less than  $ 0.002(1+z)$ \citep{Euclid_2011} and Rubin aims for bias less than $0.003(1+z)$ \citep{LSST_2019}. }  In Table \ref{tab:mean_shift}, we show bias in the mean redshift:
\beq
e_z = \frac{  \bar{z}_{ \hat{n} }  -  \bar{z}_{n_{ \rm true }}  }{ 1 +  \bar{z}_{n_{ \rm true }} }, 
\eeq
where $ \bar{z}_{ \hat{n} } $ ($\bar{z}_{n_{ \rm true }}$) denotes the mean redshift computed using the true-$z$ distribution from the estimator (direct measurement).

For SC the biases  are about 0.5\%, and are positive for all the tomographic bins.  The CZ results show larger variation and the bias goes from positive to negative as the redshift increases. The patterns for the three mass samples are similar.  Although the first bin is relatively inaccurate, the others are  good. In fact, the bias in the third bin is the smallest among all the entries. But this could be a particularity of the mock as the bias remains  0.05\% even for the 0.5\% sample.  In contrast, the SC+CZ results are stable across the tomographic bins with bias about 0.3\% for most of the entries.   Although SC+CZ bias values are systematically lower than SC in all the tomographic bins, somewhat surprisingly they are larger than CZ for the last three bins.  Nonetheless, if we take the bias of all the bins into account by adding up the absolute value of the biases in all five bins, SC+CZ gives a smaller total bias. For example, for the 1\% sample, the total bias is reduced by about 31\% and 48\% relative to the CZ and SC bias, respectively.  It seems that adding SC to CZ spoils some of the ``accidentally'' accurate bin results for CZ, but it makes the overall results more stable.   Here we avoid overinterpreting the results, and leave it to future mock tests to settle the some of the subtle trends.

%%%%%%%%%%%%%%%%%%%%%%%%%%%%%%%%%%%%%%%%%%%%%%%%%%%%%%%%%%%%%%%%%%%%%%%%%%
\begin{figure*}
    \centering
    \begin{minipage}{0.48\linewidth}
     \includegraphics[width=\linewidth]{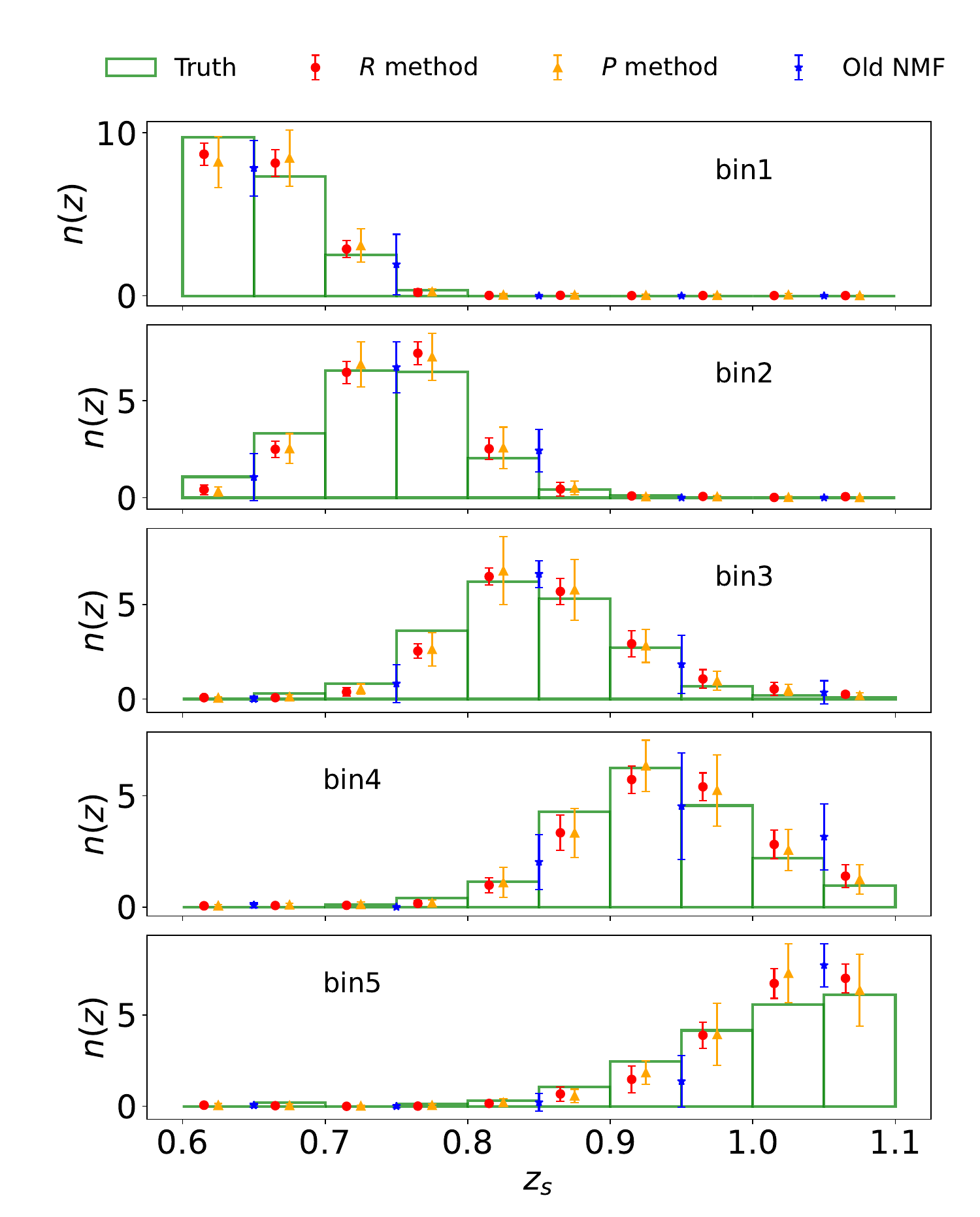}
     \caption{  Comparison of the true-$z$ distribution obtained by different implementations of the SC method with the direct measurement on the mock (green bars). The results obtained using the $R$ method (red), $P$ method (orange), and the Old NMF (blue) are compared. Because the Old NMF is not stable when it is run with resolution $\Delta z = 0.05$, we can only produce the results with  $\Delta z = 0.1$.       }
     \label{fig:histogram_sc_xu}
    % \vspace{63pt}
    \end{minipage}
    \begin{minipage}{0.01\linewidth}
       \hspace{0.01\linewidth}
    \end{minipage}
    \begin{minipage}{0.48\linewidth}
    \includegraphics[width=1\linewidth]{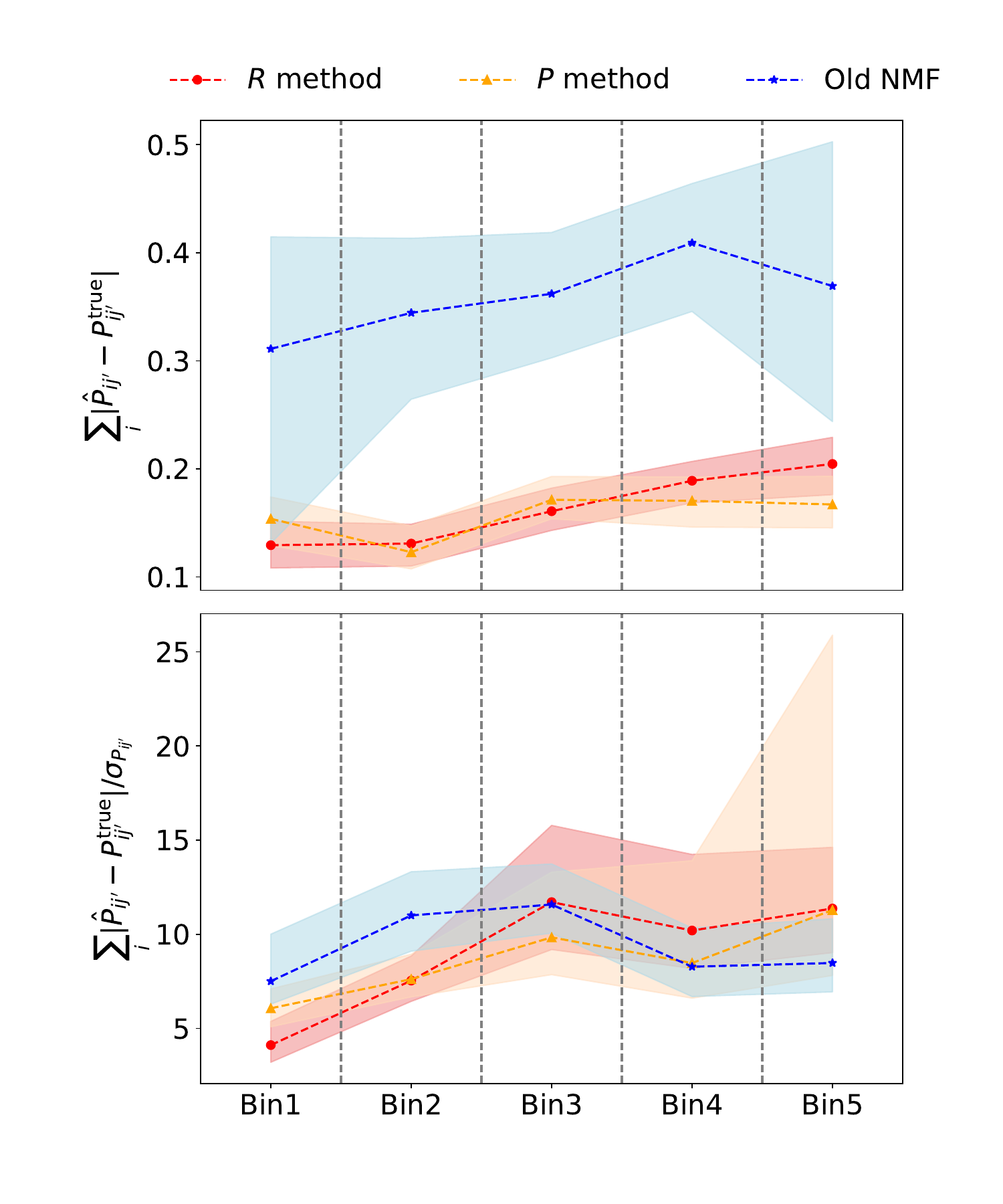}
    \caption{ Comparison of the accuracy of the estimated true-$z$ distribution obtained with different SC algorithms.  {\it Upper panel}:  $ \sum_i |\hat{P}_{ij'} - P^{\rm true}_{ij'} | $, the absolute difference between the true-$z$ distribution measured from the mock, $P^{\rm true}_{ij'} $  and the estimated one, $ \hat{P}_{ij'} $. {\it Lower panel}: $ \sum_i |\hat{P}_{ij'} - P^{\rm true}_{ij'} | / \sigma_{P_{ij'}}$ the absolute difference normalized by the estimated error  $\sigma_{P_{ij'}}$.  In both panels, the results from $R$ method (red), $P$ method (orange), and Old NMF (blue) are compared. The line and color band represent the median and the 16 and 84 percentiles among 100 mock runs.  The results for five tomographic bins are shown.  }
    \label{fig:two_piece_sc_P_R_Xu}      
    %\vspace{63pt}
    \end{minipage}
\end{figure*}

%% \begin{figure}[htb!] 
%%      \includegraphics[width=\linewidth]{histogram_sc_xu.pdf}
%%      \caption{  Comparison of the true-$z$ distribution obtained by different implementations of the SC method with the direct measurement on the mock (green bars). The results obtained using the $R$ method (red), $P$ method (orange), and the Old NMF (blue). Because the Old NMF is not stable when it is run with resolution $\Delta z = 0.05$, we can only produce the results with  $\Delta z = 0.1$.       }
%%      \label{fig:histogram_sc_xu}
%% \end{figure} 

%% \begin{figure}[htb!] 
%%     \includegraphics[width=1\linewidth]{two_piece_sc_P_R_Xu.pdf}
%%     \caption{ Comparison of the accuracy of the estimated true-$z$ distribution obtained with different algorithms for SC.  {\it Upper panel}:  $ \sum_i |\hat{P}_{ij'} - P^{\rm true}_{ij'} | $, the absolute difference between the true-$z$ distribution measured from the mock, $P^{\rm true}_{ij'} $  and the estimated one, $ \hat{P}_{ij'} $. {\it Lower panel}: $ \sum_i |\hat{P}_{ij'} - P^{\rm true}_{ij'} | / \sigma_{P_{ij'}}$ the absolute difference normalized by the estimated error  $\sigma_{P_{ij'}}$.  In both panels, the results from $R$ method (red), $P$ method (orange), and Old NMF (blue) are compared. The line and color band represent the median and the 16 and 84 percentiles among 100 mock runs.  The results for five tomographic bins are shown.  }
%%     \label{fig:two_piece_sc_P_R_Xu}
%% \end{figure} 

%% \begin{figure*}[htb!] 
%%      \includegraphics[width=\linewidth]{5bins/b/b_from_photobin_to_specbin.pdf}
%%      \caption{  \KCC{ Each subplot should be a spec-$z$ bin, and the markers should correspond to different photo-$z$ bin results.       } }
%%      \label{fig:bias_universal_leakage}
%% \end{figure*} 

\subsection{ Comparison of different implementations }

In this subsection, we compare the results obtained with different implementations.
Using SC as an example, in Fig.~\ref{fig:histogram_sc_xu}, we compare the true-$z$ distribution obtained with different algorithms for a single mock. In Fig.~\ref{fig:two_piece_sc_P_R_Xu}, we plot the absolute error and the normalized absolute error of these estimators computed with the 100-mock ensemble.

Recall that for SC, our fiducial method is the $P$ method, for which both $P_{ij'}$ and $C_{ii}' $ are the fitting parameters. Alternatively, in the $R$ method, the only unknown is  $R_{ij'}$. For the best fit value, $P$ and $R$ method give pretty similar results, but $R$ method tends to  give smaller error estimates thanks to less unknowns.

It seems that $R$ method is more constraining, but we find that it gives too small error bar too often in the case of SC+CZ. On the other hand,  $P$ method also tends to give too small error bars in the case of CZ. Because CZ is a linear problem, the computation of error bars is straightforward. We find that $R$ method yields error bars consistent with the simple error propagation result. These considerations motivate the adoption of $P$ method for SC and SC+CZ, while  $R$ method for CZ.

We further contrast our results against the one obtained with the Old NMF method, for which we use the default setting in  \cite{XU2023}.    In particular the best fit is estimated  using 100 random runs, and the best fit and the $1\sigma$ error bar are the median and the half width of the 68 percentile about the median. The stopping criterion is based on minimal $\chi^2 $ although its update rule is still rooted in the old $\mathcal{J} $. For each random run, a Gaussian perturbation is added to the measurement by sampling the covariance. We note that the covariance adopted here is derived from  100 mocks rather than the  the jackknife covariance as in  \cite{XU2023}.   The adoption of the mock covariance here is driven by the observation that there is a significant difference between the mock covariance and the jackknife covariance computed using the method in \cite{XU2023}.  \change{ By definition, the covariance is estimated by the ensemble mean among realizations; thus the mock covariance is the correct one to use. }  Furthermore, the negative measurements are set to a small positive value.  We tried to run the old NMF with $ \Delta z =0.05 $, but it is not stable in this case and we have to settle with resolution $ \Delta z = 0.1 $. \cite{Song_etal2024} also found that the old NMF fails to produce the true-$z$ distribution with a fine resolution, and used a coarse resolution of $\Delta z = 0.1$ although that did not impair the final BAO measurement there.

Fig.~\ref{fig:histogram_sc_xu}  and the top panel of Fig.~\ref{fig:two_piece_sc_P_R_Xu} demonstrate that our improved algorithm results in a much more accurate estimation of the true-$z$ distribution than the Old NMF method. The Old NMF tends to give excessively large error bars in the central region of the true-$z$ distribution, while too small error estimate in the tail. We note that when the jackknife covariance is used instead, the error bars would be smaller. %However, the mock covariance is the correct one to use.     

Table \ref{tab:mean_shift} also displays the bias from the old NMF method, and we find that our updated algorithm reduces the bias by more than a factor of 2 in almost  all the bins.

\subsection{ Estimation of the clustering amplitudes  }

Although our primary target is the true-$z$ distribution $P_{ij'}$, the accuracy of the best fit correlation function serves an important cross-check and it reflects the consistency of the algorithm.

In Fig.~\ref{fig:bias}, we plot the galaxy bias parameter of the photo-$z$ sample in the spec-$z$ bin $i$,  $ b_i^\prime $ in 10 spec-$z$ bins. Here we illustrate the results from SC+CZ using the massive 1\% sample.  The bias $b_i^\prime$ is estimated by the following means:
\begin{itemize}
\item  From the best fit $ C_{ii}' $, we have $ b_i' =\sqrt{   \frac{  C_{ii}' }{  C_{ii}^{\rm m} } }  $. 
\item  Using the best fit $C_{ii}^{\rm x} $, we get $ b_i' =  \frac{  C_{ii}^{\rm x} }{ b_i C_{ii}^{\rm m} }  $, where $ b_i $ is measured using the auto angular correlation function of the spec-$z$ sample. 
\item  Under the universal leakage assumption and  no bias evolution approximation, $ b' \approx \sum_i R_{ij'} $, which predicts that the bias is constant across all the photo-$z$ bins.  %But from Fig.~\ref{fig:bias}, there is clear variation over redshift.
 If we employ the generalized bias form,  $  b_{j'} \approx  \sum_i b_{ij'} P_{ij'} $, the resultant bias is redshift dependent albeit on index $j'$. Because the true-$z$ distribution peaks about the photo-$z$ estimate, $b_{j'} \approx b_j $. In Fig.~\ref{fig:bias}, we have plotted $b_{j'}$ as black dashed line.
\item By dividing the whole photo-$z$ sample into ten spec-$z$ bins, we can  measure their correlation functions to directly get    $b_i' $. 
\end{itemize}
Different estimates are in agreement with each other within 10\% in most of the bins in the range  $ \theta \lesssim 1^{\circ}$.  The estimate from $ C_{ii}^{\rm x}$ has larger error bar because it also requires the estimation of $b_i$  from the angular correlation function of the spec-$z$ sample. Moreover, $b_i' $ from $R_{ij'} $ is almost constant across all the spec-$z$ bins. It is remarkable that it is in nice agreement with other estimates except for the last bin  given there are a few approximations made.

%For the last two redshift bins, the estimated biases show stronger scale dependence, the agreement with the  $R_{ij'}$ estimate is limited to  $ \theta \lesssim  0.5^{\circ}$. This may be because the comoving scale is mapped to a smaller angular scale as redshift increases.  

In principle, to check the universal leakage assumption, for data in each photo-$z$ bin $j'$, we could divide the sample into 10 true-$z$ bins and measure the bias  $ b_{ij'} $. Under the universal leakage assumption, we anticipate   $ b_{ij'} $  to be independent of $j'$. 
%To check the validity of the universal leakage assumption, for data in each photo-$z$ bin $j'$, we divide the sample into five true-$z$ bins and measure the bias  $ b_{ij'} $. Under the universal leakage assumption, we have  $ b_{ij'} $  is independent of $j'$. 
However, in creating the mocks, photo-$z$'s are assigned to galaxies based on the spec-$z$ and photo-$z$ probability distribution only, without the photometry information. Thus the universal leakage assumption is built in during mock construction. %In this case it should be universal. So this is not really a test of universality assumption. Nonetheless, It serves as a background noise test. 

\begin{figure}[htb!] 
  \includegraphics[width=1\linewidth]{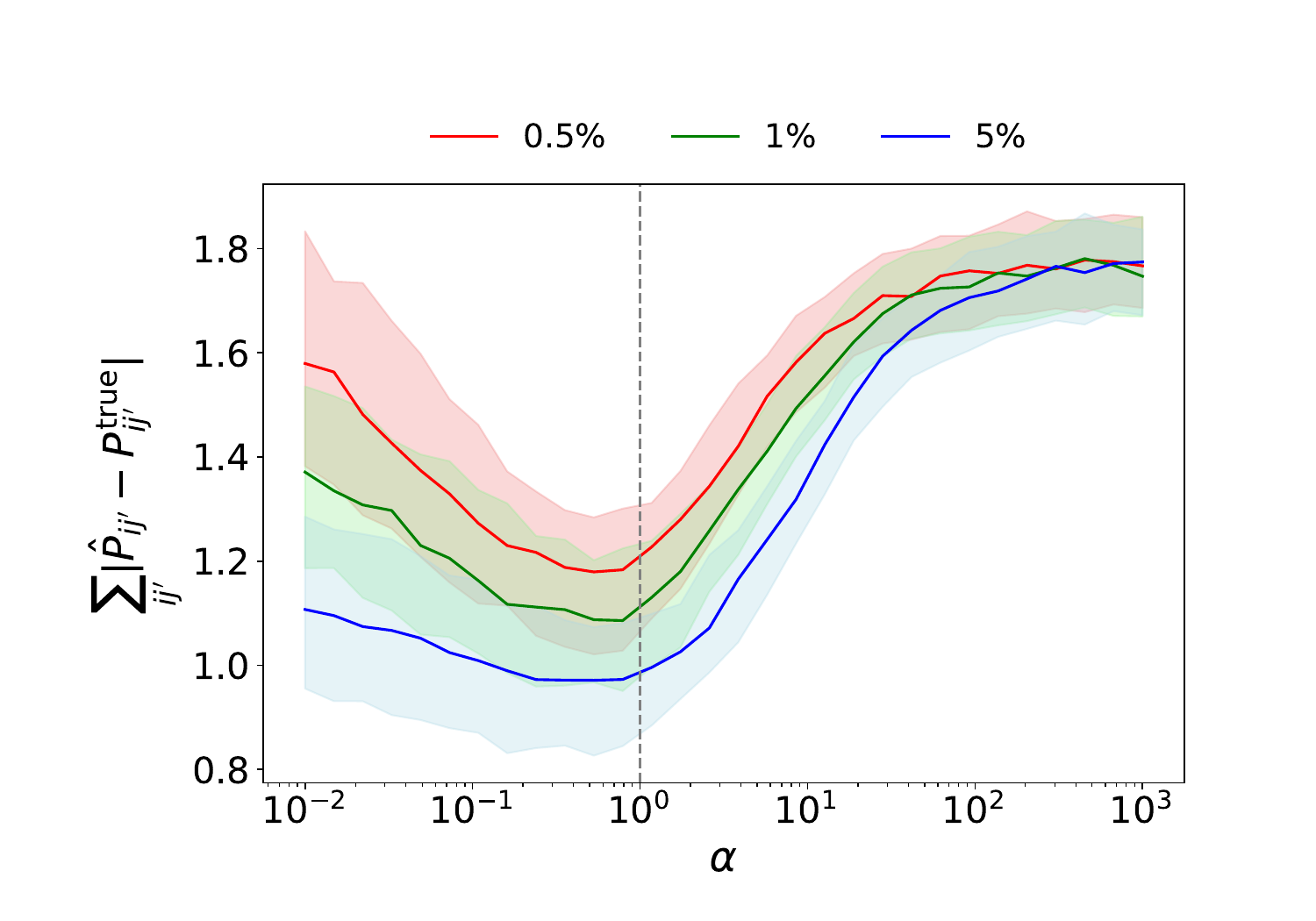}
    \caption{ Accuracy of the true-$z$ estimation characterized by  $ \sum_{ij'} | \hat{P}_{ij'} - P^{\rm true}_{ij'} |$, is plotted as a function of  $\alpha$, which measures the relative importance between SC and CZ.  The line and the associated color band represent the median and the 16 and 84 percentiles among 100 mocks. We have shown the results for three spec-$z$ sample consisting of  the most massive 5\%, 1\%, and 0.5\% galaxies in the spec-$z$ sample.  The black dashed line indicates the fiducial value $\alpha=1$.   }
    \label{fig:alpha_test_fullbins}
\end{figure}

%% \begin{figure*}
%%     \centering
%%     \begin{minipage}{0.48\linewidth}
%%     \includegraphics[width=1\linewidth]{no_bin910.pdf}
%%     % \vspace{63pt}
%%     \end{minipage}
%%     \begin{minipage}{0.01\linewidth}
%%        \hspace{0.01\linewidth}
%%     \end{minipage}
%%     \begin{minipage}{0.48\linewidth}
%%     \includegraphics[width=1\linewidth]{no_bin78910.pdf}
%%     \end{minipage}
%%     \caption{    In the absence of spec-$z$ data from the 9th and 10th bin, SC and CZ are used to jointly infer the true-$z$ distribution.  We have contrasted the cases with  $\alpha_1 = 0.1$, 1, and 10, which controls the importance of the first 8 SC bins.  Shown also are the spec-$z$ samples consisting of the most massive 5\%, 1\%, and 0.5\%.  $ \sum_{ij'} | \hat{P}_{ij'} - P^{\rm true}_{ij'} |$ is  the total absolute difference between the direct measurement and the estimated value as a function of  $\alpha_2$, which adjusts the weight of the 9th and 10th SC bin information.  } 
%%     \label{fig:a1_a2change_missingbins}    
%% \end{figure*}

\begin{figure*}[htb!] 
  \includegraphics[width=1\linewidth]{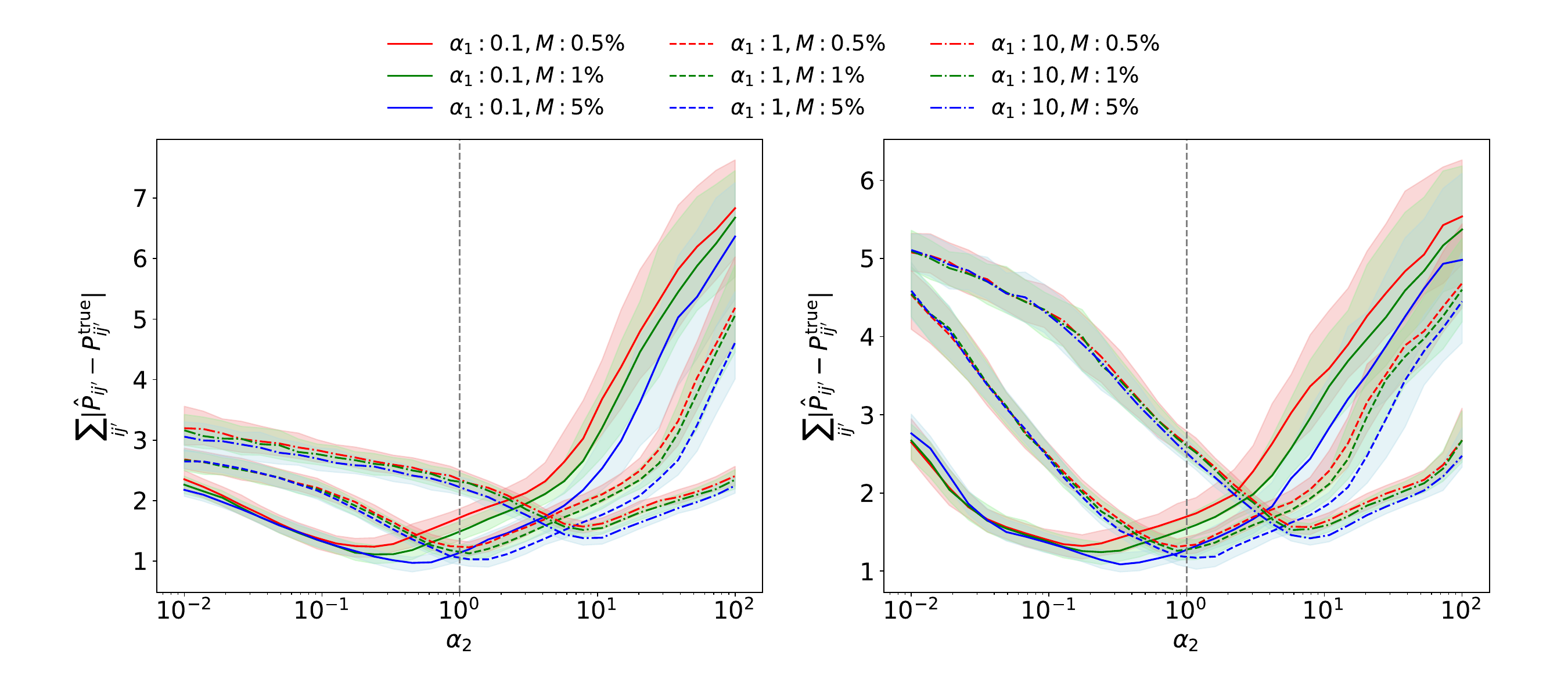}
  \caption{ Accuracy of the true-$z$ inference by SC+CZ in the absence of the high $z$ spec-$z$ data.   In the left panel, the spec-$z$ data for the last two bins, bins 9 and 10, are missing;  In the right panel, the spec-$z$ data from bins 7 to 10 are missing.    We have contrasted the cases with  $\alpha_1 = 0.1$, 1, and 10, which controls the importance of the preceding SC bins with CZ counterpart.  The total absolute difference is plotted as a function of  $\alpha_2$, which adjusts the weight of the SC bin without CZ counterpart.  Shown are the results from spec-$z$ samples consisting of the most massive 5\%, 1\%, and 0.5\%, respectively. }
    \label{fig:a1_a2change_missingbins}
\end{figure*}

\subsection{ Optimal weighting for SC and CZ } 

While $\sigma^2 $ takes into account the measurement error of the correlation function, the theoretical uncertainties of the method are not included.  Simply adding up $\mathcal{J}_1 $ and  $\mathcal{J}_2 $ implicitly assumes that they have equal constraining power and this may not lead to optimal results because SC and CS have different level of degeneracy. A simple way to tackle this issue is to assign different weights to their cost functions. We consider two scenarios: (i) the number of spec-$z$ bins with spec-$z$ data is equal to the number of photo-$z$ bins or (ii) the number of spec-$z$ bins with spec-$z$ data is less than the number of photo-$z$ bins. The second setup is particularly interesting because the lack of spec-$z$ data at high $z$ limits the application of the clustering-based method to the high $z$ regime.

\subsubsection{ Full spec-$z$ data }

Here we focus on the scenario when the number of spec-$z$ bin is the same as the number of photo-$z$ bins. For this full spec-$z$ bin data case, we consider the joint cost function:
\beq
\mathcal{J} = \alpha \mathcal{J}_1 + \mathcal{J}_2 , 
\eeq
where $ \mathcal{J}_1$ and  $\mathcal{J}_2 $ are given by  Eqs.~\eqref{eq:J1_cost} and \eqref{eq:J2_cost} respectively, and  the weight $\alpha $ adjusts their relative importance.

In Fig.~\ref{fig:alpha_test_fullbins}, we show the absolute error of the estimated distribution. We note that we have summed over both indices $i$ and $j'$. We have shown the results for three spec-$z$ samples, which respectively consist of the most massive 5\%, 1\%, and 0.5\% of the galaxies. The best fit and the $1\sigma$  error band are derived from 100 mocks.   For  $\alpha \gtrsim 100$, we see that the curves are convergent because the information is dictated by SC in this limit.   On the other hand, in the low $\alpha$ limit, the result is determined by CZ.  As expected, if the fraction of galaxies in the spec-$z$ sample is higher,  the signal-to-noise of the cross correlation function is higher, and hence the absolute error is smaller.

\change{ In the intermediate range, for the 0.5\% sample, there is a dip roughly around $\alpha \sim 0.7  $, implying that this weight optimally combines the information.  As the mass fraction increases, dip becomes much broader and shallower but the dip position remains nearly unchanged with very mild shift towards a smaller $\alpha$.  These indicate  that the SC information helps little and the signal is largely dominated by CZ.    We note that the  from 0.5\% to 1\%, the optimal $\alpha $ seems to increase, but this is not expected and should be interpreted as statistical fluctuations.  }  For the 1\% spec-$z$ sample, the optimal combination of SC+CZ  reduces the total error by 20\% relative to CZ and 40\% relative to SC.

\begin{figure*}
    \centering
    \begin{minipage}{0.48\linewidth}
     \includegraphics[width=\linewidth]{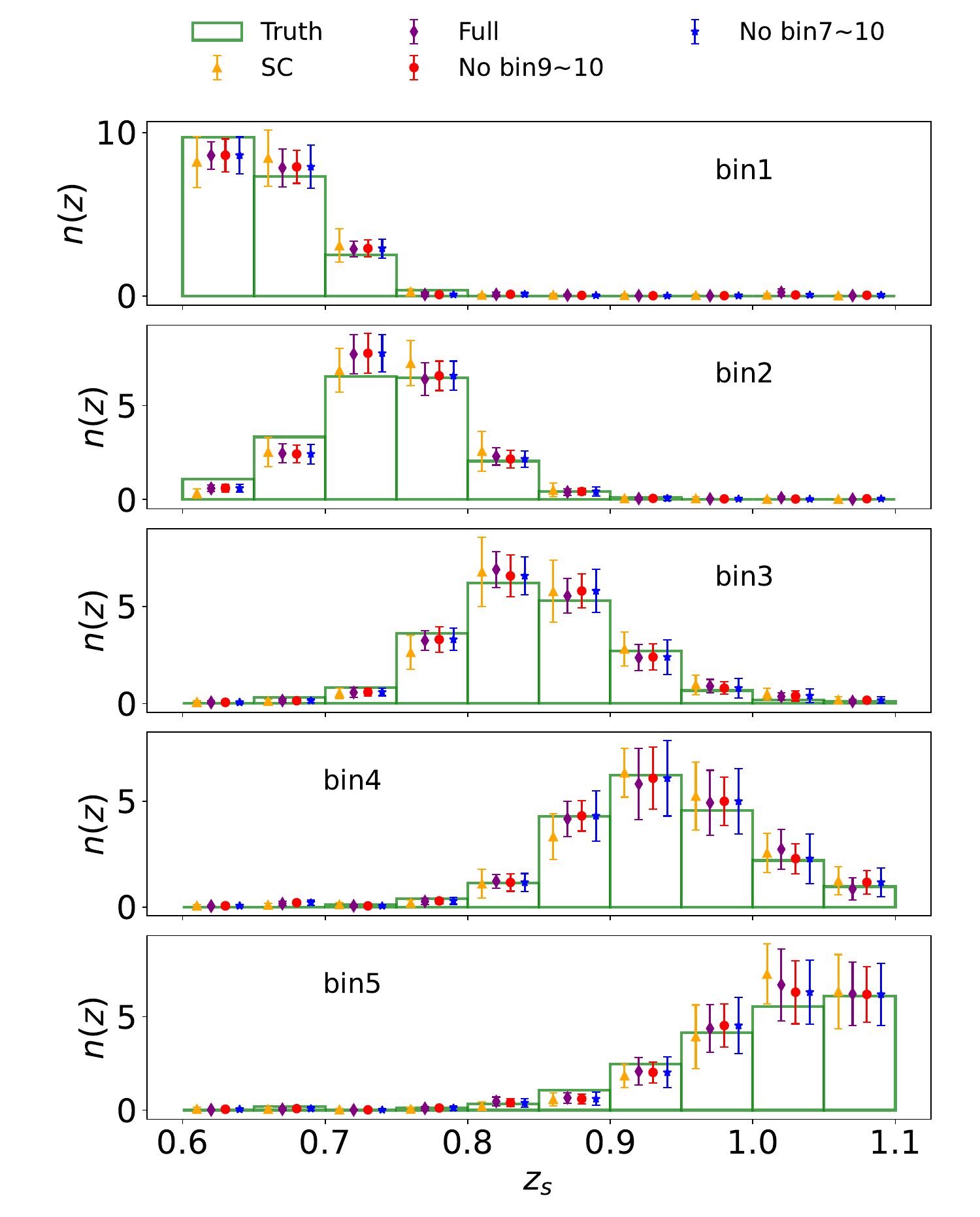}
     \caption{  Comparison of the true-$z$ distribution obtained by different number of spec-$z$ bins with spec-$z$ data available against the direct measurement on the mock (green bars) for a single mock.   We have presented the results from full spec-$z$ bins (violet),  no spec-$z$ bin 9 and 10 (red), no spec-$z$ bin 7 to 10 (blue), and SC only (orange). } % With the reduction of the number of spec-$z$ bins, the constraining power is weakened.  }
     \label{fig:histogram_miss_bin_logM_1p}
    % \vspace{63pt}
    \end{minipage}
    \begin{minipage}{0.01\linewidth}
       \hspace{0.01\linewidth}
    \end{minipage}
    \begin{minipage}{0.48\linewidth}
     \includegraphics[width=\linewidth]{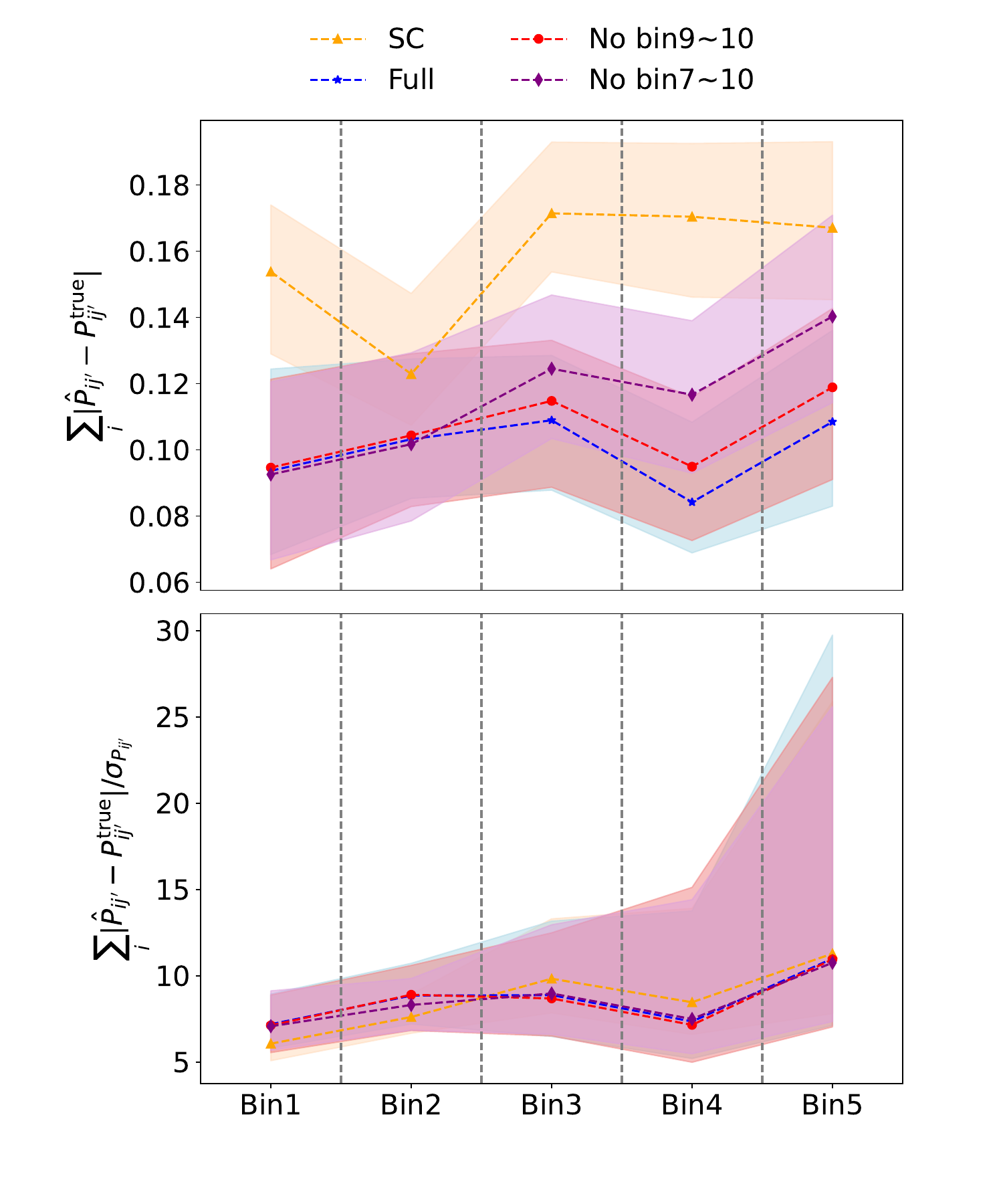}
     \caption{  The absolute error of the true-$z$ distribution estimated by different amount of spec-$z$ bin data.
  {\it Upper panel}:  $ \sum_i |\hat{P}_{ij'} - P^{\rm true}_{ij'} | $, the absolute difference between the true-$z$ distribution measured from the mock, $P^{\rm true}_{ij'} $  and the one estimated, $ \hat{P}_{ij'} $. {\it Lower panel}: $ \sum_i |\hat{P}_{ij'} - P^{\rm true}_{ij'} | / \sigma_{P_{ij'}}$ the absolute difference normalized by the estimated error  $\sigma_{P_{ij'}}$.  In both panels, the results from full spec-$z$ data (blue), no spec-$z$ bin 9 and 10 (red), no spec-$z$ bin 7 to 10 (violet), and SC only (orange) are compared.  With the reduction of the number of spec-$z$ bin data, the constraining power is weakened. When only the high-$z$ spec-$z$ bins are missing, the impact is mainly localized in the high-$z$ tomographic bins with the low-$z$ bins little affected. }
     \label{fig:two_piece_miss_bin_logM_1p}
    \end{minipage}
\end{figure*}

\subsubsection{ Missing spec-$z$ data }

Often the spec-$z$ data are only available in the relatively low redshift range.  Thus it is of interest to consider the situation in which the low $z$ part of the true-$z$ distribution is simultaneously constrained by both SC and CZ, while the high $z$ part relies solely on SC.

%% \begin{align}
%%       \mathcal{J}_{1} & =\frac{1}{2} \sum_{\theta,i^{'},j^{'}}^{i^{'}=4,j^{'}=4}\big[ D_{i^{'}j^{'}}(\theta)-\sum_{k}P_{ki^{'}}P_{kj^{'}}C_{kk}(\theta)  \big]^{2}  \\
%%      &  +\alpha\frac{1}{2}\sum^{j^{'}=5}_{\theta,j^{'}}\big[ D_{5j^{'}}(\theta)-\sum_{k}P_{k5}P_{kj^{'}}C_{kk}(\theta)  \big]^{2} \\

%%       &+\alpha\frac{1}{2}\sum^{i^{'}=5}_{\theta,i^{'}}\big[ D_{i^{'}5}(\theta)-\sum_{k}P_{ki^{'}}P_{k5}C_{kk}(\theta)\big]^{2}  \\
%%      & -\alpha\frac{1}{2}\big[ D_{55}(\theta)-\sum_{k}P_{k5}P_{k5}C_{kk}(\theta)\big]^{2}
%% \end{align}

For notational convenience, let $S$ be the set of all the bins of interest, $ \{1, 2, \dots, N \} $, $T $ be the set of the bins with spec-$z$ data available, $\{1,2,,\dots, N_s  \}$ with $N_s \le N$,  and $S \backslash T$ be the set of bins without spec-$z$ info, $\{N_s+1, \dots, N \}$.   $T \bigotimes T$  denotes the tuple  $\{ (i,j)| \, i\in T \textrm{ and } j \in T\} $, and  its complement $ (T \bigotimes T)^C $ represents $\{ (i,j)| \, i\in S \backslash T \textrm{ or } j \in S \backslash T\} $.    With these notations defined, the total cost function with two weight adjustment factors $\alpha_1 $ and $\alpha_2$ is written as 
\begin{equation}
    \mathcal{J}=  \mathcal{J}_{1}+\mathcal{J}_{2},
\end{equation}
where $\mathcal{J}_1$ is given by
\begin{align}
  \mathcal{J}_{1}   & = \frac{\alpha_1}{2} \sum_{  (i^{'}, j^{'}) \in  T  \bigotimes T  } \sum_\mu F_{i'j'\mu}  \nn \\
          &  +    \frac{ \alpha_2 }{2} \sum_{    (i^{'}, j^{'} ) \in  (T \bigotimes T)^C   } \sum_\mu   F_{i'j'\mu}, 
\end{align}
with $ F_{i'j'\mu} $ denoting 
\begin{align}
F_{i'j'\mu} = \frac{ W_1(\theta_\mu)  }{ \sigma^2_{i' j'}(\theta_\mu) }   \big[ D_{i^{'}j^{'}}(\theta_\mu ) - \sum_{k}P_{ki^{'}}P_{kj^{'}}C_{kk}'( \theta_\mu )  \big]^{2},
 \end{align}
%%   & 2 \mathcal{J}_{1}  =  \sum_{ \substack{ \mu \\ (i^{'}, j^{'}) \in  T  \bigotimes T}  } \frac{ W_1(\theta_\mu)  }{ \sigma^2_{i' j'}(\theta_\mu) }  \big[ D_{i^{'}j^{'}}(\theta_\mu ) - \sum_{k}P_{ki^{'}}P_{kj^{'}}C_{kk}( \theta_\mu )  \big]^{2}   \nn  \\
%%   &  +   \alpha_2  \sum_{   \substack{  \mu \\
%%     (i^{'}, j^{'} ) \in  (T \bigotimes T)^C  }  }  
%%   \frac{ W_1(\theta_\mu )}{  \sigma^2_{i' j'}(\theta_\mu) }   
%%   \big[ D_{i^{'}j^{'}}(\theta)-\sum_{k}P_{ki^{'}}P_{kj^{'}}C_{kk}(\theta)  \big]^{2}  . 
%% \end{align}
and  $\mathcal{J}_2 $ by 
\beq
    \mathcal{J}_{2} =\frac{1}{2}  \sum_{(i,j^{'}) \in T \bigotimes T} \sum_\mu  \frac{ W_2(\theta_\mu) }{ \sigma_{ij'}^2 (\theta_\mu) }  \Big[ D_{ij^{'}}(\theta_\mu )-P_{ij^{'}}C_{ii}^{\rm x}(\theta_\mu)  \Big]^{2} .  
\eeq
In words, the weight $\alpha_1$ is to adjust the importance of the SC part with respect to  the CZ counterpart, while the additional weight $\alpha_2$ in  $\mathcal{J}_1 $  \change{ allows for the possibility to }  leverage the weight of the SC contribution from the $S \backslash T$ bins to compensate the missing CZ contribution. 

% \begin{align}
%   &  \qquad \qquad  2 \mathcal{J}_{1} =  \nn \\
%   &   \sum_{ \substack{ \mu \\ i^{'} \in  T,  j^{'} \in  T }  } \frac{ W_1(\theta_\mu)  }{ \sigma^2_{i' j'}(\theta_\mu) }  \big[ D_{i^{'}j^{'}}(\theta_\mu ) - \sum_{k}P_{ki^{'}}P_{kj^{'}}C_{kk}( \theta_\mu )  \big]^{2}  \nn  \\
%   &  +   \alpha_2  \sum_{  \substack{ \mu \\
%   \{   i^{'} \in  T,  j^{'} \in  S \backslash T  \}   \\  \cup  \, \{  i^{'} \in  S \backslash T,  j^{'} \in  T \} }  }  
%   \frac{ W_1(\theta_\mu )}{  \sigma^2_{i' j'}(\theta_\mu) }   
%   \big[ D_{i^{'}j^{'}}(\theta)-\sum_{k}P_{ki^{'}}P_{kj^{'}}C_{kk}(\theta)  \big]^{2}  . 
% \end{align}

In the left panel of Fig.~\ref{fig:a1_a2change_missingbins}, we show the results when the ninth and tenth spec-$z$ bin data are missing.  We contrasted the cases with $ \alpha_1 $ fixed to be 0.1, 1, and 10, corresponding to the CZ information being dominant, similar to, and subdominant to the matching SC bins. The accuracy of the inference is presented as a function of   $ \alpha_2 $.   In addition, we have compared the results from the spec-$z$ samples consisting of the most massive 5\%, 1\%, and 0.5\% of the galaxies.

First, for $\alpha_1 = 0.1 $, CZ information is determinant in the first 8 bins.  For large $ \alpha_2$, the importance of SC bin 9 and 10 is inflated and becomes dominant so that the overall accuracy decreases.  The error is minimized at a trough varying from $\alpha_2 \sim 0.2  $ to 0.5, depending on the fraction of the spec-$z$ sample.  The optimal $ \alpha_2 $ is less than unity reflects that the constraining power of the SC in 9th and 10th bin is  bin-wisely  weaker than CZ in the bins 1 to 8.  The accuracy also deteriorates in the limit of small $ \alpha_2 $  because the missing bins become less and less constrained in this case.
%We also see that the higher mass fraction spec-$z$ sample yields a slightly stronger results, and this results from the complementary albeit mild constraint from the lower $z$ bins with CZ information.
The optimal $\alpha_2$ for higher mass sample is slightly higher in order to balance the greater CZ constraint from the lower redshift bins.  The differences among the three spec-$z$ samples  diminish  at  $\alpha_2 \lesssim 0.1 $, and this  reflects that the CZ  information is saturated and the missing bin is the bottleneck in enhancing the accuracy of the true-$z$ inference.

The overall trend with $\alpha_2$ is similar for $\alpha_1 =1$  and 10, but the shape generally shifts to a larger $\alpha_2 $ to be in line with $\alpha_1 $.   For $ \alpha_1 = 1 $, the CZ and SC information has comparable weight for the first 8 bins, we find that  the minimal error occurs around $ \alpha_2 \sim 1  $ for 0.5\% spec-$z$ sample and  $ \sim 2  $ for 5\%.  The differences among the spec-$z$ samples are less pronounced than  $\alpha_1=0.1 $ case because larger $\alpha_1  $ value  reduces the weight of CZ.  Finally, when SC information is dominant in the first 8 bins ($\alpha_1 = 10 $), the optimal  $\alpha_2$  moves to $\sim  9$ because the constraining power of SC bins are roughly similar.

\change{ It is worth summarizing the roles of the weight factors.  $ \alpha_1 $ adjusts the weight of the SC bins relative to their CZ counterpart. The SC and CZ part compete with each other and they have equal weights when $\alpha_1 = 1 $.  $\alpha_2 $ controls the weight of the SC part without CZ information.  Through the above exercise, we see that  $\alpha_2 $ is in the same order of magnitude  as the dominating part of the lower redshift bins.  }

In addition, we present  the missing spec-$z$ data in bin 7-10 case  in the right panel of Fig.~\ref{fig:a1_a2change_missingbins}.  They are qualitatively similar to the no bin 9 and 10  case. In particular, the optimal $\alpha_2 $  is quite similar to the no bin 9 and 10 case. %, but we also notice a small shift towards smaller $\alpha_2$ value among all $\alpha_1$ cases.  From Fig.~\ref{fig:two_piece_sc_cz_sc+cz_logM_1p}, we see that the amount of increase in  absolute error is stronger in CZ than the reduction in SC from tomographic bin 4 to 5. Thus a smaller $\alpha_2 $ can reduce the importance
However, in contrast, the minimal error increases in the CZ dominating case ($ \alpha_1 = 0.1 $) by a small amount, while the SC dominated case ($ \alpha_1 = 10 $) is the least affected. This can be attributed to the fact that the CZ information is more constraining than SC.   For the same reason, in the small $ \alpha_2$  limit,  we find that the accuracy is generally lower than the no bin 9 and 10 case.

%With more spec-$z$ bins missing,  the optimal $ \alpha_2 $ shifts to a larger value to enhance the importance of the SC bins. Again it is more apparent in the $ \alpha_1 = 0.1 $ case, with the SC dominating case barely affected as the last two bins are the least constraining ones among the SC bins.  

Although the precise weights depend on the relative constraining power of the SC and CZ data, our test suggests that there is generally a set of optimal $\alpha_1$ and $\alpha_2 $ minimizing the error.

We show the bias in the mean redshift in Table \ref{tab:mean_shift} for missing spec-$z$ bin data 9-10 and 7-10. Even with missing spec-$z$ bin data, SC+CZ is stable and the results are similar to the full bin case. Sometimes, the mean value in missing bin case is even lower than the full bin case although the difference is statistically insignificant.  Only when the bin 7-10 are missing, we notice systematic increase in bias in the tomographic bin 4 and 5 (by about half $\sigma$).

%% \begin{figure}[tb!] 
%%     \includegraphics[width=1\linewidth]{5bins/alpha/a1_1_a2_change_only_top.pdf}
%%     \caption{    In the absence of spec-$z$ data from the 5th bin, SC and CZ are used to jointly infer the true-$z$ distribution.  We have contrasted the cases with  $\alpha_1 = 0.1$, 1, and 10, which controls the importance of the first four SC bins.  Shown also are the spec-$z$ samples consisting of the most massive 5\%, 1\%, and 0.5\%.  $ \sum_{ij'} | \hat{P}_{ij'} - P^{\rm true}_{ij'} |$, the total absolute difference between the direct measurement and the estimated value as a function of  $\alpha_2$, which adjusts the weight of the 5th SC bin information.  } 
%%     \label{fig:a1_1_a2_change}
%% \end{figure} 

%% \begin{figure}[tb!] 
%%     \includegraphics[width=1\linewidth]{5bins/alpha/missbin45_a1_a2_change_only_top.pdf}
%%     \caption{ Similar to Fig.~\ref{fig:a1_1_a2_change}, except  for the case without spec-$z$ data from bin 4 and 5. }
%%  %     \KCC{ Remove the lower panel, can't get much out of it.}   }
%%     \label{fig:a1_1_a2_change_No4_5}
%% \end{figure} 

Finally, for completeness, we plot the true-$z$ distribution estimated by various number of spec-$z$ bin data  for a \change{single} mock in Fig.~\ref{fig:histogram_miss_bin_logM_1p} and the absolute error estimated from 100 mocks in Fig.~\ref{fig:two_piece_miss_bin_logM_1p}.  In these plots,  we assume $\alpha_1= 1 $ and $ \alpha_2 = 1 $, which are close \change{to}  the optimal weights for the 1\% sample used here. 
We have shown the results obtained with full spec-$z$ bins, no spec-$z$ bin 9 and 10, no spec-$z$ bin 7 to 10, and SC only. We indeed see that as the number of spec-$z$ bins decreases, the constraining power weakens. The true-$z$ distribution for lower tomographic bins 1 and 2 are almost unaffected for the missing spec-$z$ bin cases, while the higher tomographic bin ones are more affected. %With less spec-$z$ bins, the estimated error bar sizes tend to become larger.
In summary, we have demonstrated that by incorporating the SC information with CZ, we are able to extend the constraint on the true-$z$ distribution to higher redshift where spec-$z$ data are missing, and the constraint is better \change{than} SC alone.

\section{Conclusions} 
\label{sec:conclusions}

Characterization of the true-$z$ distribution is crucial in the science of the wide-field imaging surveys such as weak lensing and BAO.  A valuable avenue to calibrate the true-$z$ distribution of a photo-$z$ sample is to make use of the clustering information. Clustering-$z$ (CZ) has been widely used in the cosmological analyses in imaging surveys. However, a limitation of this method is that the spec-$z$ sample is often only available in relatively low redshift regime.  The self-calibration (SC) method relying solely on the clustering information of the photometric sample is gaining popularity.  In this work we develop a method to jointly constrain the true-$z$  distribution of a photo-$z$ sample using the information of SC and CZ simultaneously.  We use the DES Y3 catalog to test our method and find that it can effectively infer the true-$z$ distribution. The codes performing the SC+CZ inference can be downloaded on GitHub\footnote{ \url{https://github.com/kcc274/SCplusCZ} }.

Previous SC analyses often rely on the non-negative matrix factorization (NMF) algorithm.  Such an interpretation becomes a burden when we want to generalize the method to combine SC with CZ. Inspired by the  NMF method, we construct multiplicative update rules to directly solve for the true-$z$ distribution. It is worth stressing that we have avoided the NMF interpretation altogether. Our straightforward analysis enables us to easily combine the information of SC with CZ.

Our formalism has improved upon the previous approach by taking the weighting functions into account. We have included the inverse error weighting and an additional weighting function. The inverse error weighting allows us to downweight the impact of the poor measurements and upweight the good ones.  The additional weighting function give the freedom to put more weight on the constraining part of the correlation function, and we take it to be the form $ \theta^{-n} $ with $ n = 1$.  We demonstrate that our improved algorithm results in a much more accurate estimation of the true-$z$ distribution compared to the previous method in the case of SC (Figs.~\ref{fig:histogram_sc_xu} and \ref{fig:two_piece_sc_P_R_Xu}). Moreover, our algorithm gives more stable results and allows us to have a true-$z$ distribution with higher resolution.  This bias in the mean redshift is reduced by more than a factor of two.

We employ our algorithm to show that SC+CZ  improves the constraint relative to using SC or CZ alone (Figs.~\ref{fig:histogram_sc_cz_sc+cz_logM_1p} and \ref{fig:two_piece_sc_cz_sc+cz_logM_1p}).  SC+CZ  gives  a  stable bias on the mean redshift, keeping it at a level of $\sim 0.3\%$, even though it sometimes fairs worst than CZ.  The clustering-based methods also yield estimates of the intrinsic clustering amplitude. We find that various clustering measurements are consistent with each other (Fig.~\ref{fig:bias}).  To optimize the constraining power of SC and CZ data, we assign extra weight factors to the SC cost function.  We consider the scenario with  the number of spec-$z$ bins with spec-$z$ data  equal to the number of photo-$z$ bins and the more interesting one with the spec-$z$ bin data only available in the low $z$ bins. We find that there generally exist some weights that minimize the total error (Figs.~\ref{fig:alpha_test_fullbins} and \ref{fig:a1_a2change_missingbins}). The precise weights depend on the constraining power of the SC and CZ data, and detailed mock tests are required to locate the optimal values. To highlight the improvement, we quote the 1\% spec-$z$ sample result with full spec-$z$ bins here. In this case we find that  the optimal combination reduces  the total error by 20\%  and 40\%  compared to using respectively CZ and SC only.  The test with spec-$z$ data restricted to the low $z$ range demonstrates that by incorporating with the SC information, we can extend the utility of the clustering-based method to higher redshift.

After successfully demonstrating the power of the SC+CZ method via mock catalogs, it is desirable to apply it to real data, such as the DES Y6 BAO data \citep{DESY6_BAO,DESY6_BAOsample}. For this dataset,  only the portion of data with $z<1$ is calibrated with CZ, while the data in the range $1< z<1.2 $ have to be calibrated with a spec-$z$ sample from a small sky area.  A few caveats need to keep in mind when applying the method to real data.   \change{ In this work we have used  $ \Delta z = 0.05 $, but this is still larger than the redshift bin width often taken in CZ analysis (e.g.~$ \Delta z =  0.03 $). We have also assumed no bias evolution, at least in the CZ case.  We leave it to future work to test the impact of these setups, and study possible improvements.  We have not considered the magnification bias in this paper because the redshift range of the data is still small and the mock does not include this effect. For larger redshift extent, magnification bias correction must be made.  Furthermore, the photometric data are more prone to observational systematics, which must be treated before SC.  Compared to the applications to the clustering samples such as the lens sample in weak lensing and the BAO sample mentioned above,  the application of the SC method to the source sample in weak lensing is more challenging  simply because it is not constructed for clustering analysis. The source sample is often inhomogeneous and the random catalog is likely missing in standard analysis, so it must be built with great care.  Moreover, the systematics mitigation is designed for the shear measurements, and so  it remains to show that the existing mitigation efforts are sufficient or additional clustering weight needs to be created. }  Nevertheless, our work paves the way to calibrate the true-$z$ distribution in high redshift using the clustering information.

%The universal leakage assumption has to be tested. 

\begin{acknowledgements}
  \change{ We thank the anonymous referee for his/her insightful comments that improve the presentation of the manuscript. }  WZ, KCC, and RS are  supported by the National Science Foundation of China under the grant number 12273121 and  the science research grants from the China Manned Space Project with NO.CMS-CSST-2021-B01.   HX is supported by the National SKA Program of China (grant No. 2020SKA0110100), the National Natural Science Foundation of China (Nos. 11922305, 11833005) and the science research grants from the China Manned Space Project with NOs. CMS-CSST-2021-A02. LZ is supported by National SKA Program of China (2020SKA0110401, 2020SKA0110402, 2020SKA0110100), the National Key R\&D Program of China (2020YFC2201600), the China Manned Space Project with No. CMS-CSST-2021 (A02, A03), and Guangdong Basic and Applied Basic Research Foundation (2024A1515012309).
\end{acknowledgements}

\bibliographystyle{aa}
\bibliography{references}

%------------------------------------------------------------------------------%
%------------------------------------------------------------------------------%
\begin{appendix}

\section{Derivation of the multiplicative update rule  }  
\label{Appendix:Derivation_update_rule}

In general, a multiplicative update rule is a set of iterative instructions to update a function by multiplying it with some factor to search for the minimum of the cost function.   The most influential multiplicative update rule for non-negative matrix factorization (NMF) is presented by \cite{LeeSeung_2000}. This multiplicative update rule can be interpreted as a special kind of gradient descent rule with variable step size.   A distinguishing feature of the NMF update rule is the non-negativity of the solution. This has been adopted in \cite{Zhang_etal2017} to solve the SC equation.  The update rule in \cite{LeeSeung_2000} can be easily derived by splitting the gradient of the cost function into the positive and negative parts \citep{Choi_2008}. Inspired by this derivation, we derive multiplicative update rules for the joint inference by SC and CZ.

%We here directly compute the update rule for $P$. However, a caveat is that while  $\mathcal{J}_2 $  is convex for $P$,  $\mathcal{J}_1 $ it is not.  The consequences need to be checked numerically. 

%% For SC+CZ, we consider the cost function $\mathcal{J}$ defined as 
%% \begin{equation}
%%   \label{eq:J1_J2_cost}
%%     \mathcal{J}=\mathcal{J}_{1}+\mathcal{J}_{2},
%% \end{equation}
%% where $\mathcal{J}_{1}$ and $\mathcal{J}_{2}$ are the contributions from self-calibration and clustering-$z$ to $\mathcal{J}$, respectively.  They are given by 
%% \begin{align}
%%     \mathcal{J}_{1} & =\frac{1}{2} \sum_{\theta,i^{'},j^{'}}\big[ D_{i^{'}j^{'}}(\theta)-\sum_{k}P_{ki^{'}}P_{kj^{'}}C_{kk}(\theta)  \big]^{2} ,\\
%%      \mathcal{J}_{2} & =\frac{1}{2} \sum_{\theta,i,j^{'}} \big[ D_{ij^{'}}(\theta)-P_{ij^{'}}C_{ii}^{\rm x}(\theta)  \big]^{2} ,  
%% \end{align}
%% where $D$ denotes the data measurement (angular correlation function in our case), and $C_{ii}^{\rm x} $ is the diagonal spec-$z$ correlation function, and $P_{ij'}$  is the scattering rate from the $i$ spec-$z$ bin to $j'$ photo-$z$ bin.

\subsection{ $P$ method  }

In $P$ method, we aim to derive multiplicative rules for  $ P_{ij^{'}} $,  $C_{ii}^{\rm x} $, and  $C'_{ii} $ so that the cost function $\mathcal{J} $ [Eq.~\eqref{eq:J1_J2_cost}] is minimized.  If we take $P_{ab^{'}}$ as the independent variable, we also need to account for  the normalization constraint Eq.~\eqref{eq:P_normalization}. This can be implemented by introducing the Lagrange multipliers into the cost function as
\begin{equation}
\label{eq:J1_J2_Pconstraint}
    \mathcal{J}=\mathcal{J}_{1}+\mathcal{J}_{2}-\sum_{j^{'}}\lambda_{j^{'}}\big(\sum_{i}P_{ij^{'}}-1 \big). 
\end{equation}
We  first computed the derivative with respect to $P_{ab^{'} }$ and then to the Lagrange multipliers.

The derivative of $\mathcal{J}_1 $ with respect to $P_{ab^{'} }$ reads 
\begin{align}
    %% \frac{\partial \mathcal{J}_{1}}{\partial P_{ab^{'}}}&=\sum_{\theta,i^{'},j^{'}} \big(D_{i^{'}j^{'}}(\theta)-\sum_{k}P_{ki^{'}}P_{kj^{'}}C_{kk}(\theta )\big)  \big[-\sum_{k}\frac{\partial}{\partial P_{ab^{'}}} \big(P_{ki^{'}}P_{kj^{'}}C_{kk}(\theta)  \big) \big] \nn \\
    %% &=\sum_{\theta,i^{'},j^{'}} \big(D_{i^{'}j^{'}} (\theta)-\sum_{k}P_{ki^{'}}P_{kj^{'}}C_{kk}(\theta)  \big) \big(-\sum_{k}\delta_{ka}\delta_{i^{'}b^{'}}P_{kj^{'}}C_{kk}(\theta)   \big) \times2 \nn \\
    %% &=-2\sum_{\theta,i^{'},j^{'}}(D_{i^{'}j^{'}}(\theta)-\sum_{k}P_{ki^{'}}P_{kj^{'}}C_{kk}(\theta )) \delta_{i^{'}b^{'}}P_{aj^{'}}C_{aa}(\theta) \nn \\
  \frac{\partial \mathcal{J}_{1}}{\partial P_{ab^{'}}}   = -2\sum_{j^{'}, \mu }  & \frac{ W_1(\theta_\mu)  }{\sigma_{b' j'}^2( \theta_\mu )  }  \Big[ D_{b^{'}j^{'}}(\theta_\mu )-\sum_{k}P_{kb^{'}}P_{kj^{'}}C'_{kk}(\theta_\mu ) \Big]      \nn \\
  &  \times   P_{aj^{'}}C'_{aa} (\theta_\mu ) ,
\end{align}
and for $\mathcal{J}_{2}$, we have
\begin{align}
    %% \frac{\partial \mathcal{J}_{2}}{\partial P_{ab^{'}}} &=\sum_{\theta,i,j^{'}} \big(D_{ij^{'}}(\theta)-P_{ij^{'}}C_{ii}^{\rm x} (\theta)\big)\big[-\frac{\partial}{\partial P_{ab^{'}}}(P_{ij^{'}}C_{ii}^{\rm x}(\theta) )\big] \nn \\
    %% &=\sum_{\theta,i,j^{'}} \big(D_{ij^{'}}(\theta)-P_{ij^{'}}C_{ii}^{\rm x} (\theta) \big)(-\delta_{ai}\delta_{b^{'}j^{'}}C_{ii}^{\rm x}(\theta)) \nn \\
\frac{\partial \mathcal{J}_{2}}{\partial P_{ab^{'}}}  &=- \sum_\mu \frac{ W_2(\theta_\mu) }{ \sigma_{ab'}^2(\theta_\mu) }  [ D_{ab^{'}}(\theta_\mu )-P_{ab^{'}}C_{aa}^{\rm x}(\theta_\mu )]C_{aa}^{\rm x}(\theta_\mu ).
\end{align}

Without the constraint, following \cite{LeeSeung_2000}, the update rule is given by
\begin{align}
\label{eq:Pabp_update_nonorm}
P_{ab^{'}}\xleftarrow{}P_{ab^{'}}\frac{ [\partial_{P_{ab'}} \mathcal{J}]^-}{[\partial_{P_{ab'}} \mathcal{J}]^+},
\end{align}
where $[\partial_{P_{ab'}} \mathcal{J}]^-$ ($[\partial_{P_{ab'}} \mathcal{J}]^+$) denotes the unsigned negative (positive) part of the derivative of $\partial_{P_{ab'}} \mathcal{J} $. We can contrast this with the usual additive gradient descent \citep{GoodBengCour16}. The term  $[\partial_{P_{ab'}} \mathcal{J}]^-$ tends to increase the value of $  P_{ab^{'} } $, while $[\partial_{P_{ab'}} \mathcal{J}]^+$ tends to decrease it, so they play the same role as in the usual gradient descent. When the minimum is reached, $\nabla \mathcal{J} $ vanishes and the multiplicative factor becomes unity.  Moreover, the factor  $ [\partial_{P_{ab'}} \mathcal{J}]^- / [\partial_{P_{ab'}} \mathcal{J}]^+ $ is non-negative, and so it does not flip the sign of  $ P_{ab^{'}} $.

When the normalization constraint is included, we have 
\begin{align}
 \frac{\partial \mathcal{J}}{\partial P_{ab^{'}}}&=\frac{\partial \mathcal{J}_{1}}{\partial P_{ab^{'}}}+\frac{\partial \mathcal{J}_{2}}{\partial P_{ab^{'}}}-\lambda_{b^{'}}.
\end{align}
For the treatment of the Lagrange multiplier part, we follow the procedures in \cite{ZhuYangOja_2013}. We can first establish a preliminary update rule in terms of $\lambda_{b^{'}}$, and then demand the resultant $P_{ab^{'}}$  to satisfy the normalization constraint. This enabled us to solve for $\lambda_{b^{'}}$. It is convenient to include a negative sign for $\lambda_{b^{'}}$ and assume $\lambda_{b^{'}}$to be positive. However, this assumption does not always hold and so \cite{ZhuYangOja_2013} consider a ``moving term'' trick to alleviate this problem.

 The preliminary update rule is
\begin{align}
P_{ab^{'}}\xleftarrow{}P_{ab^{'}}\frac{ [\partial_{P_{ab'}} \mathcal{J}]^- +\lambda_{b^{'}}}{[\partial_{P_{ab'}} \mathcal{J}]^+}.
\end{align}
The new $P$ is also demanded to meet the normalization constraint: 
 \begin{align}
\sum_{a} & P_{ab^{'}}\frac{ [\partial_{P_{ab' }} \mathcal{J}]^- +\lambda_{b^{'}}}{[\partial_{P_{ab'}} \mathcal{J}]^+ }=1 . %, \\
%\lambda_{b^{'}}&=\frac{1-\sum_{a}P_{ab^{'}}\frac{[\partial_{P_{ab'}} \mathcal{J}]^-}{ [ \partial_{P_{ab'}} \mathcal{J}]^+ }}{ \sum_{a}\frac{P_{ab^{'}}}{[\partial_{P_{ab'}} \mathcal{J}]^+}}.
\end{align}
Solving for $\lambda_{b'}  $, the update rule becomes
\begin{align}
P_{ab^{'}}\xleftarrow{}P_{ab^{'}}  \frac{ [\partial_{P_{ab'} }\mathcal{J}]^- \sum_{c}\frac{P_{cb^{'}}}{ [\partial_{P_{cb'} }\mathcal{J}]^+ }+1-\sum_{c}P_{cb^{'}}\frac{ [\partial_{P_{cb'} }\mathcal{J}]^- }{ [\partial_{P_{cb'} }\mathcal{J}]^+}}   {[\partial_{P_{ab'} }\mathcal{J}]^+\sum_{c}\frac{P_{cb^{'}}}{ [\partial_{P_{cb'} }\mathcal{J}]^+ }}.
\end{align}
To make the numerator always positive, we applied the ``moving trick'' to get
\begin{align}
    \label{eq:update_rule_Pijp}
P_{ab^{'}}\xleftarrow{}P_{ab^{'}}\frac{ [\partial_{P_{ab'} }\mathcal{J}]^- \sum_{c}\frac{P_{cb^{'}}}{ [\partial_{P_{cb'} }\mathcal{J}]^+ }+1}   {[\partial_{P_{ab'} }\mathcal{J}]^+ \sum_{c}\frac{P_{cb^{'}}}{ [\partial_{P_{cb'} }\mathcal{J}]^+}+\sum_{c}P_{cb^{'}}\frac{[\partial_{P_{cb'} }\mathcal{J}]^-}{[\partial_{P_{cb'} }\mathcal{J}]^+}}.
\end{align}

We also update $C_{aa}'$ [in Eq.~\eqref{eq:J1_cost}] and  $C_{aa}^{\rm x} $ [in Eq.~\eqref{eq:J2_cost}]  using \change{the} multiplicative rule although this is not necessary. Because they are linear parameters, direct analytic  minimization is possible as was done in \cite{Zhang_etal2017}.
The update rule for $C_{aa}'$ reads 
\begin{align}
  \label{eq:update_rule_Cii}
C_{aa}'(\theta)\xleftarrow{}C_{aa}'(\theta) \frac{ [\partial_{C'_{aa}(\theta) } \mathcal{J} ]^-  }{ [\partial_{C'_{aa}(\theta) } \mathcal{J} ]^+  }.
\end{align}
\change{A} similar update rule for  $C_{aa}^{\rm x}$ can be derived.

\subsection{ $ R $ method }

In $R$ method, only $R_{ij'} $ is the unknown. Explicitly the cost functions read 
\begin{align}
  \label{eq:J1_R_cost}  
    \mathcal{J}_{1} & =\frac{1}{2} \sum_{i^{'},j^{'}, \mu }  \frac{ W_1(\theta_\mu) }{ \sigma_{i'j'}^2 (\theta_\mu) }  \Big[ D_{i^{'}j^{'}}(\theta_\mu) - \sum_k R_{ki'} R_{kj'}  C_{kk}^{\rm m} (\theta_\mu)  \Big]^{2} ,\\
  \label{eq:J2_R_cost}  
    \mathcal{J}_{2} & =\frac{1}{2} \sum_{i,j^{'}, \mu}  \frac{ W_2(\theta_\mu) }{ \sigma_{ij'}^2 (\theta_\mu) }  \Big[ D_{ij^{'}}(\theta_\mu ) -  R_{ij'} b_i C_{ii}^{\rm m} (\theta_\mu)  \Big]^{2} .  
\end{align}

The derivative of the cost functions with respect to $R_{ab'}$ are given by  
\begin{align}
    %% \frac{\partial \mathcal{J}_{1}}{\partial P_{ab^{'}}}&=\sum_{\theta,i^{'},j^{'}} \big(D_{i^{'}j^{'}}(\theta)-\sum_{k}P_{ki^{'}}P_{kj^{'}}C_{kk}(\theta )\big)  \big[-\sum_{k}\frac{\partial}{\partial P_{ab^{'}}} \big(P_{ki^{'}}P_{kj^{'}}C_{kk}(\theta)  \big) \big] \nn \\
    %% &=\sum_{\theta,i^{'},j^{'}} \big(D_{i^{'}j^{'}} (\theta)-\sum_{k}P_{ki^{'}}P_{kj^{'}}C_{kk}(\theta)  \big) \big(-\sum_{k}\delta_{ka}\delta_{i^{'}b^{'}}P_{kj^{'}}C_{kk}(\theta)   \big) \times2 \nn \\
    %% &=-2\sum_{\theta,i^{'},j^{'}}(D_{i^{'}j^{'}}(\theta)-\sum_{k}P_{ki^{'}}P_{kj^{'}}C_{kk}(\theta )) \delta_{i^{'}b^{'}}P_{aj^{'}}C_{aa}(\theta) \nn \\
  \frac{\partial \mathcal{J}_{1}}{\partial R_{ab^{'}}}   = -2\sum_{j^{'}, \mu }  & \frac{ W_1(\theta_\mu)  }{\sigma_{b' j'}^2( \theta_\mu )  }  \Big[ D_{b^{'}j^{'}}(\theta_\mu )- \sum_{k}R_{kb^{'}}R_{kj^{'}}C^{\rm m}_{kk}(\theta_\mu ) \Big]      \nn \\
  &  \times   R_{aj^{'}}C^{\rm m}_{aa} (\theta_\mu ) ,
\end{align}
and 
\begin{align}
    %% \frac{\partial \mathcal{J}_{2}}{\partial P_{ab^{'}}} &=\sum_{\theta,i,j^{'}} \big(D_{ij^{'}}(\theta)-P_{ij^{'}}C_{ii}^{\rm x} (\theta)\big)\big[-\frac{\partial}{\partial P_{ab^{'}}}(P_{ij^{'}}C_{ii}^{\rm x}(\theta) )\big] \nn \\
    %% &=\sum_{\theta,i,j^{'}} \big(D_{ij^{'}}(\theta)-P_{ij^{'}}C_{ii}^{\rm x} (\theta) \big)(-\delta_{ai}\delta_{b^{'}j^{'}}C_{ii}^{\rm x}(\theta)) \nn \\
\frac{\partial \mathcal{J}_{2}}{\partial R_{ab^{'}}}  &=- \sum_\mu \frac{ W_2(\theta_\mu) }{ \sigma_{ab'}^2(\theta_\mu) }  [ D_{ab^{'}}(\theta_\mu )- b_a R_{ab^{'}}C_{aa}^{\rm m}(\theta_\mu )] b_a C_{aa}^{\rm m}(\theta_\mu ).
\end{align}
The update rule for $ R_{ab'} $ can then be constructed: 
\begin{align}
\label{eq:Pabp_update_nonorm}
R_{ab^{'}}\xleftarrow{} R_{ab^{'}}\frac{ [\partial_{R_{ab'}} \mathcal{J}]^-}{[\partial_{R_{ab'}} \mathcal{J}]^+}.
\end{align}
We note that $ P_{ij'} $ follows from Eq.~\eqref{eq:Pijp_noevolution} if we assume that there is no galaxy bias evolution.

\section{ Impact of outlying galaxies  }
\label{sec:Impact_OutlyingGal}

When the photo-$z$ galaxies are present only in the redshift range $[0.6,1.1]$, the true-$z$ distribution of this sample can extend outside it. In this work, we restrict both photo-$z$ and spec-$z$ range of the galaxies to [0.6,1.1], and this effectively assumes that $P_{ij'} $ vanishes outside this range. In practice, however, this kind of cleaning is not possible.

%Since we enforce that $P_{ij'} $ is normalized to unity, the missing correlation does not affect the normalization. Nonetheless this still affect the absolute photo-$z$ and spec-$z$ clustering. For example, the photo-$z$ bias parameters may disagree slightly from the one computed using the cut-off $ P_{ij]} $. In the main text, the default sample is obtained by removing the galaxies with spec-$z$ lying outside the range [0.6,1.1].
%The cutoff may be unavoidable, especially at high redshift. However, in practice, cleaning is not possible.
% In this appendix, we check the impact of these outlying galaxies.  Consequently, this impact of outlying galaxies is minimal for our sample. 

\begin{figure}[htb!] 
    \includegraphics[width=1\linewidth]{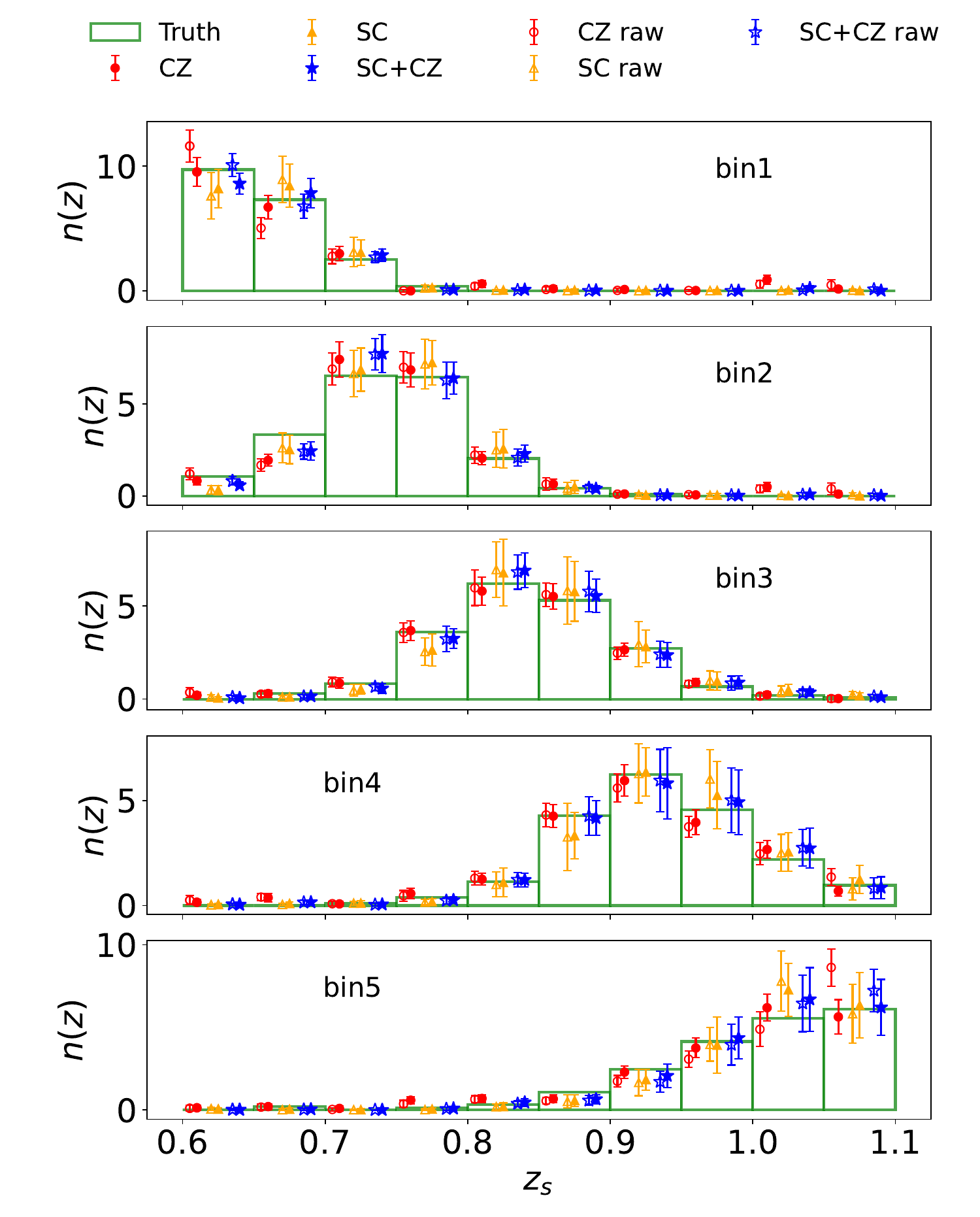}
    \caption{ Test of the impact of the galaxies with spec-$z$ values lying outside the redshift range [0.6,1.1]. The true-$z$ distribution from the raw sample, which contains the outlying spec-$z$ galaxies (unfilled markers), are compared with the  cleaned sample results (filled markers). The results from SC (red), CZ (orange), and SC+CZ (blue) are displayed. The direct mock measurements (green bars) are from the cleaned sample.       }
    \label{fig:histogram_sc_cz_sc+cz_logM_1p_no_default_compare}
\end{figure}

To test the impact of the outlying galaxies on the estimation of the true-$z$ distribution, we apply the algorithm to the raw photo-$z$ sample, which contains the galaxies with spec-$z$ values lying outside the range [0.6,1.1]. When the outlying galaxies are not removed, this additional component does not correlate with the ones inside, and this causes a reduction in the correlation signals in general.  We compare the resulting distributions from the raw sample against the ones from the cleaned sample in Fig.~\ref{fig:histogram_sc_cz_sc+cz_logM_1p_no_default_compare}.   Overall, we find that the impact of the outlying galaxies is pretty small for our sample.

We note that, as expected, the distributions for the boundary bins are more affected, i.e.~bin 1 and 5. This is because the outlying spec-$z$ galaxies are mainly located in the photo-$z$ bins near the boundaries. We further  note that CZ is more strongly affected than SC, and thus SC+CZ is in between. This can be explained by the fact that in CZ, the correlation is localized in the precise redshift of the spec-$z$ bin, while the information in SC is more distributed, and hence CZ is more affected by the reduction in correlation with the boundary bins.

%To be concrete, let us consider the correlation for the first spec-$z$ bin. For in the first photo-$z$ bin, when the spec-$z$ data outside the range [0.6,1.1] is removed, the correlation between the external spec-$z$ bin 1 data is increased because after removal there is a larger fraction of spec-$z$ galaxies in the first photo-$z$ bin. For 

\section{Impact of setting negative measurements to a tiny positive value }
\label{sec:Impact_NoNegative}

In the main text, we mention that the multiplicative rule implicitly assumes that the measurements are positive. A quick fix is to set the negative measurements to a tiny positive value, for which we take it to be $ 10^{-5} $.   Here we test the impact of such a modification on the true-$z$ inference results.

\begin{figure*}
    \centering
    \begin{minipage}{0.48\linewidth}
     \includegraphics[width=\linewidth]{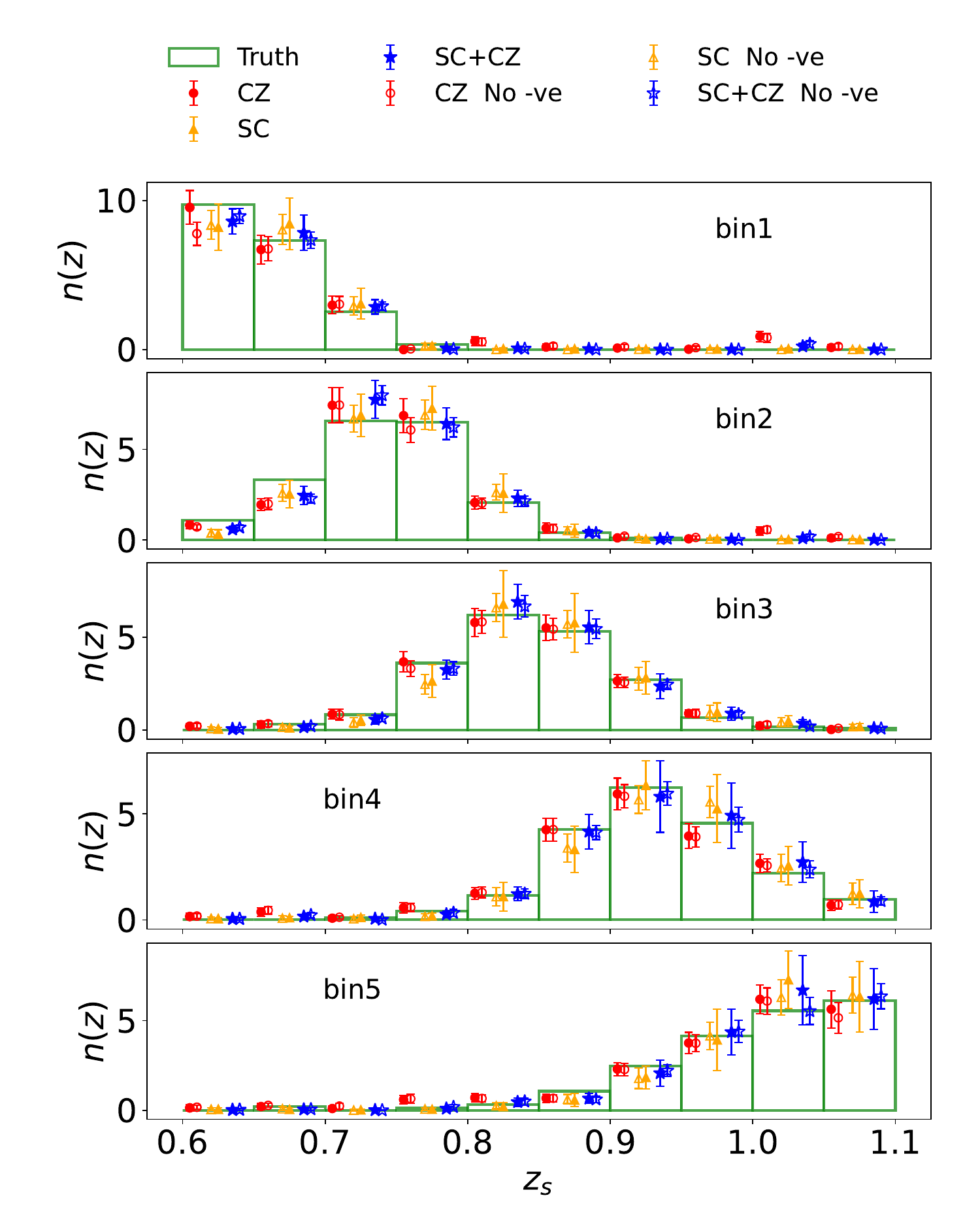}
     \caption{  Comparison of the true-$z$ distribution obtained by setting the negative measurements to a tiny positive value (empty markers) or not (filled markers).  We have presented the results from SC (orange), CZ (red), and SC+CZ (blue).    }       
     \label{fig:histogram_NoNeg_logM_1p}
    % \vspace{63pt}
    \end{minipage}
    \begin{minipage}{0.01\linewidth}
       \hspace{0.01\linewidth}
    \end{minipage}
    \begin{minipage}{0.48\linewidth}
     \includegraphics[width=\linewidth]{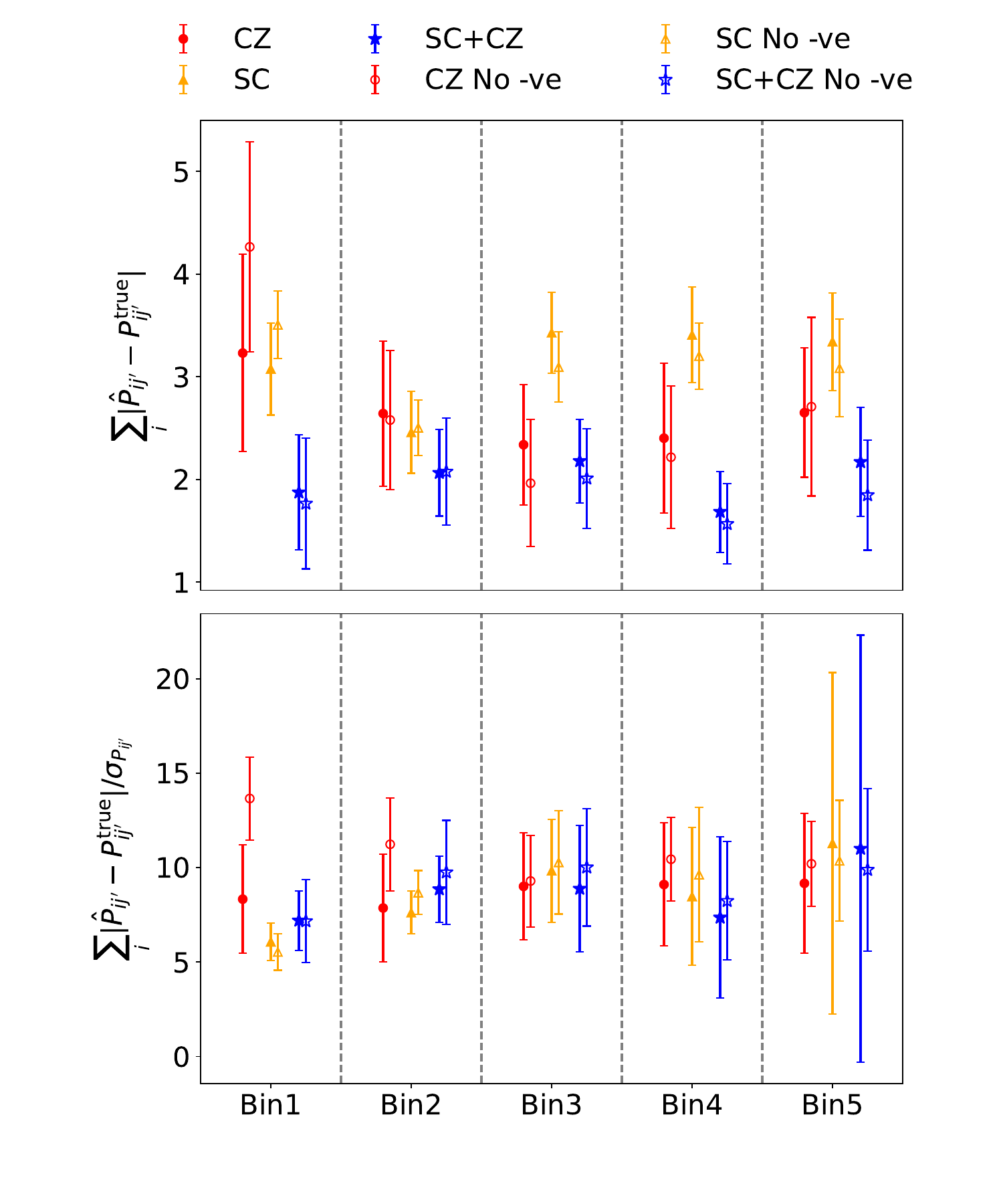}
     \caption{  Absolute error of the true-$z$ distribution estimated using samples processed by  setting the negative measurements to a tiny positive value (empty markers) or not (filled markers). Shown are the results from   SC (orange), CZ (red), and SC+CZ (blue).
   {\it Upper panel}:  $ \sum_i |\hat{P}_{ij'} - P^{\rm true}_{ij'} | $, the absolute difference between the true-$z$ distribution measured from the mock, $P^{\rm true}_{ij'} $  and the one estimated, $ \hat{P}_{ij'} $. {\it Lower panel}: $ \sum_i |\hat{P}_{ij'} - P^{\rm true}_{ij'} | / \sigma_{P_{ij'}}$ the absolute difference normalized by the estimated error  $\sigma_{P_{ij'}}$.  }
     \label{fig:two_piece_NoNeg_logM_1p}
    \end{minipage}
\end{figure*}

Fig.~\ref{fig:histogram_NoNeg_logM_1p}  showcases the true-$z$ distribution from a mock. While there are only small fluctuations in the best fit value, without the modification, the estimated error bars from SC and SC+CZ appear to be larger, but the CZ error bars are little affected.   We further examine the accuracy of the estimated distribution in Fig.~\ref{fig:two_piece_NoNeg_logM_1p}. The central value and the  $1\sigma$  error bound are estimated from 100 mock catalogs.  From the top panel, we see that nulling the negative measurements seems to increase the accuracy of SC and SC+CZ in most of the bins albeit by a statistically insignificant amount, but the modification also causes  the SC  bin 1 result to degrade.
The impact on the estimated error bar is more apparent.  The modification homogenizes  the range of $ |\hat{P}_{ij'} - P^{\rm true}_{ij'} | / \sigma_{P_{ij'}}$ across tomographic bins.

There seem to be benefits, especially on the error bar estimation, in doing such a modification, and thus we may consider adopting it in future.

\end{appendix}

\end{document}